%% file: main.tex
\documentclass[sigconf, nonacm]{acmart}
\input{packages}

\input{macro}
\AtBeginDocument{%
  }

\begin{document}
\begin{sloppypar}

\title{Are We There Yet? Unraveling the State-of-the-Art Graph Network Intrusion Detection Systems}
\author{Chenglong Wang}
\affiliation{%
  \institution{Shanghai Jiao Tong University}
  \city{Shanghai}
  \country{China}}
\email{wangchenglong25@sjtu.edu.cn}

\author{Pujia Zheng}
\affiliation{%
  \institution{Shanghai Jiao Tong University}
  \city{Shanghai}
  \country{China}}
\email{zhengpujia@sjtu.edu.cn}

\author{Jiaping Gui}
\affiliation{%
  \institution{Shanghai Jiao Tong University}
  \city{Shanghai}
  \country{China}}
\email{jgui@sjtu.edu.cn}

\author{Cunqing Hua}
\affiliation{%
  \institution{Shanghai Jiao Tong University}
  \city{Shanghai}
  \country{China}}
\email{cqhua@sjtu.edu.cn}

\author{Wajih Ul Hassan}
\affiliation{%
  \institution{University of Virginia}
  \city{Virginia}
  \country{USA}}
\email{hassan@virginia.edu}

\renewcommand{\shortauthors}{Wang et al.}

\begin{abstract}

Network Intrusion Detection Systems (NIDS) are vital for ensuring enterprise security. Recently, Graph-based NIDS (\gnids) have attracted considerable attention because of their capability to effectively capture the complex relationships within the graph structures of data communications. Despite their promise, the reproducibility and replicability of these \gnids remain largely unexplored, posing challenges for developing reliable and robust detection systems. This study bridges this gap by designing a systematic approach to evaluate state-of-the-art \gnids, which includes critically assessing, extending, and clarifying the findings of these systems. We further assess the robustness of \gnids under adversarial attacks. Evaluations were conducted on three public datasets as well as a newly collected large-scale enterprise dataset. Our findings reveal significant performance discrepancies, highlighting challenges related to dataset scale, model inputs, and implementation settings. We demonstrate difficulties in reproducing and replicating results, particularly concerning false positive rates and robustness against adversarial attacks. This work provides valuable insights and recommendations for future research, emphasizing the importance of rigorous reproduction and replication studies in developing robust and generalizable \gnids solutions.

\end{abstract}

\renewcommand\footnotetextcopyrightpermission[1]{}



%%

\maketitle

\input{tex/introduction}

\input{tex/motivation}

\input{tex/approach}
\input{tex/evaluation}
\input{tex/relatedwork}

\input{tex/conclusion}

\bibliographystyle{ACM-Reference-Format}
\bibliography{references}

\input{tex/appendix}
\end{sloppypar}
\end{document}

%% file: packages.tex
\usepackage{graphicx}
\usepackage[linesnumbered,lined,vlined,ruled,commentsnumbered,noend]{algorithm2e}
\usepackage{textcomp}
\usepackage{xcolor}
\usepackage{listings}
\usepackage{url}
\usepackage{xspace}
\usepackage{multirow}
\usepackage{balance}
\usepackage{ctable}
\usepackage{listings}
\usepackage{color}
\usepackage{graphicx}
\usepackage{subcaption}
\usepackage[nobiblatex]{xurl}

\usepackage{amsmath}
\usepackage{amsfonts}

\usepackage{algpseudocode}
\usepackage{caption}
\usepackage{subcaption}
\usepackage{hyperref}
\hypersetup{
    colorlinks=true,
    linkcolor=blue,
    citecolor=blue,
    urlcolor=blue,
}
\usepackage[nameinlink]{cleveref}
\usepackage{acronym}
\usepackage{fancybox}
\usepackage{verbatim}
\usepackage[normalem]{ulem}
\usepackage{float}
\usepackage{newfloat}
\usepackage{mdframed}
\usepackage{enumitem}
\usepackage{enumerate}
\usepackage{booktabs}
\usepackage{pifont}
\usepackage{threeparttable}
\usepackage{makecell}

\usepackage{array}
\usepackage{multirow}
\usepackage{ragged2e}
\usepackage{graphicx}
\usepackage{babel}

%% file: macro.tex
\usepackage{acronym}
\definecolor{red}{RGB}{255,0,0}
\definecolor{green}{RGB}{18,220,168}

\newcommand{\ie}{i.e.,\xspace}
\newcommand{\eg}{e.g.,\xspace}
\newcommand{\etc}{etc.\xspace}
\newcommand{\etal}{\textit{et al.}\xspace}

\newcommand{\pj}[1]{\textcolor{blue}{(PJ: #1)}}

\newcommand{\wajih}[1]{\textcolor{green}{(WAJIH: #1)}}

\newcommand{\gnids}{\textsc{GIDS}\xspace}

\newcommand{\toolname}{\textsc{GidsRep}\xspace}

\newcommand{\eat}[1]{}
\acrodef{wp}[WP]{Website Fingerprinting}
\acrodef{apt}[APT]{Advanced Persistent Threats}
\acrodef{lol}[LotL]{Living-Off-The-Land}
\acrodef{ids}[IDS]{Intrusion Detection System}
\acrodef{vae}[VAE]{Variational AutoEncoder}
\acrodef{re}[RE]{reconstruction error}
\acrodef{sv}[SV]{Stableness Value}
\acrodef{as}[AS]{anomaly score}
\acrodef{gas}[GAS]{graph anomaly score}
\acrodef{etw}[ETW]{Event Tracing for Windows}
\acrodef{e3}[E3]{Engagement 3}
\acrodef{e5}[E5]{Engagement 5}
\acrodef{ttp}[TTPs]{Tactics, Techniques, and Procedures}
\acrodef{hsg}[HSG]{High-level Scenario Graph}
\acrodef{nlp}[NLP]{Natural Language Processing}
\acrodef{dg}[DG]{Detection Graph}
\acrodef{poi}[POI]{Point of Interest}
\acrodef{iv}[IV]{Important Value}
\acrodef{sg}[SG]{Suspicious Graph}
\acrodef{mttd}[MTTD]{Mean Time to Detect}
\acrodef{soc}[SOC]{Security Operations Center}
\newlength{\MaxSizeOfLineNumbers}%
\definecolor{keywordcolor}{rgb}{0.8,0.1,0.5}
\definecolor{lightlightgray}{gray}{.96}
\definecolor{lightgray}{gray}{.925}
\definecolor{medlightgray}{gray}{0.7}
\definecolor{medgray}{gray}{0.4}
\definecolor{darkgray}{gray}{0.35}
\definecolor{nearblack}{gray}{0.15}

\newcommand{\anomale}{Anomal-E\xspace}
\newcommand{\vgrnn}{VGRNN\xspace}
\newcommand{\pikachu}{PIKACHU\xspace}
\newcommand{\euler}{EULER\xspace}
\newcommand{\argus}{ARGUS\xspace}

\usepackage{mdframed}

\newenvironment{custombox}[1]
  {\vspace{5pt} % Increase space above the box
  \begin{mdframed}[
      backgroundcolor=white,
      linecolor=black,
      linewidth=0.5pt,
      roundcorner=0pt,
      innertopmargin=5pt,
      innerbottommargin=5pt,
      innerrightmargin=5pt,
      innerleftmargin=5pt
    ]
    \textbf{#1} }
  {\end{mdframed}
  \vspace{5pt}} % Increase space below the box

  \newcommand{\PP}[1]{
\vspace{2px}
\noindent{\bf \IfEndWith{#1}{.}{#1}{#1.}}
}

%% file: tex/introduction.tex
\section{Introduction}
In the face of increasingly sophisticated cyber threats, adversaries have leveraged various vulnerabilities and zero-day exploits to infiltrate targeted systems. These attacks are often meticulously planned, stealthy, and long-term, with the primary objective of exfiltrating sensitive data from large enterprises and government agencies. Such breaches have resulted in substantial economic losses, as evidenced by high-profile incidents involving Yahoo~\cite{Yahoo}, Marriott~\cite{Marriott}, and Home Depot~\cite{HomeDepot}. Network Intrusion Detection Systems (NIDS) are indispensable in enterprise security infras\-tructures, continuously monitoring network traffic logs to identify suspicious signatures and patterns indicative of potential threats and generating alerts for security analysts.

Traditionally, existing NIDS, such as Zeek and Snort, operate using a predefined set of rules to detect anomalous behaviors~\cite{zeek, suricata, snort, modSecurity, yara}. While rule-based NIDS are known for their interpretability, they struggle to detect unknown attack patterns (high false negative rates), particularly zero-day exploits. Moreover, these systems often produce high false positive rates 
%when encountering sophisticated or slightly varied attack patterns
due to inaccurate or too generic definitions of rules, which leads to ``threat alert fatigue'', where security analysts become overloaded and unable to respond effectively~\cite{alahmadi202299,hassan2019nodoze}. This undermines the overall effectiveness of the enterprise security infrastructure.

To address these limitations, Graph-based NIDS (\gnids) have
been developed as a crucial component of enterprise network security~\cite{yu2018netwalk, paudel2022pikachu, king2023euler, kipf2016variational, xu2024understanding,cheng2021step, cai2021structural}. \gnids construct directed graphs from network communications, where nodes represent machines and edges represent communication flows. They utilize graph encoders to generate embeddings of nodes and edges based on traffic data within the network and often employ temporal encoders to capture graph temporal dynamics. By learning  traffic patterns, \gnids detect any deviations as suspicious behavior. They exhibit robust detection performance, even against unknown threats like zero-day exploits, and offer superior adaptability compared to traditional NIDS. Despite their potential, the reproducibility and replicability (R+R) of \gnids remain largely unexplored, posing significant challenges for the development of reliable and robust detection systems. These challenges encompass the absence of open-source code, experimental configurations, hyperparameter settings, as well as limitations in datasets, computational resources, efficiency, and robustness against attacks.

%\pj{On the one hand, due to the lack of open-source code by the authors of the proposed methods and issues with hyperparameter settings and experimental environment configurations, it is difficult for researchers to reproduce the experiments. On the other hand, because of the limitations of dataset size and issues related to computational resources, efficiency, and security against attacks, researchers find it difficult to apply these methods to real-world scenarios in enterprises.}

Our study aims to tackle the R+R issues in enterprise security research by systematically evaluating state-of-the-art (SOTA) \gnids. Our study seeks to confirm, question, and clarify the results of previous research, ensuring their validity and reliability across various contexts. This is crucial for developing robust and generalizable detection systems capable of effectively combating evolving security threats. We conduct a comprehensive assessment of these systems on both public datasets and a newly collected enterprise dataset, measuring their detection performance, efficiency from both temporal and spatial perspectives, and robustness under adversarial attacks.

To study existing \gnids, we delineate our research questions into three pivotal domains: Implementation, Evaluation, and Deployment.

\noindent \textbf{Implementation.} In the implementation domain, our focus is on the intricacies of \gnids re-implementations and their impact on system performance. Specifically, \emph{RQ1: What are the key factors that influence the re-implementations of \gnids?} This involves a meticulous analysis of various aspects to ensure accurate experimental settings, including the optimal tuning of hyperparameters. This domain emphasizes the necessity of understanding how different configurations and hyperparameters affect the overall effectiveness of the system, enabling us to identify the most impactful settings for developing a functional and optimal system.

%\textbf{Implementation} In the implementation domain, our focus is on the intricacies of Graph NIDS configurations and their impact on system performance. Specifically, \emph{RQ1: What key implementation parameters influence the detection efficacy of Graph NIDS?} This involves a meticulous analysis of various settings to ensure optimal tuning of these parameters, thereby enhancing model performance and robustness against cyber threats. This domain emphasizes the necessity of understanding how different configurations and hyperparameters affect the overall effectiveness of the system, enabling us to identify the most impactful settings for optimal performance. 
%\pj{the implementation domain should focus on just re-implementing prior work, i.e., by developing a functional system that orchestrates all of the various components of a (graph) NIDS. Probably, this RQ should be rewritten, or a new one should be introduced which focuses on an aspect touched in the abstract, i.e., "We demonstrate difficulties in reproducing and replicating results," this means that the authors have undergone much trouble and reimplementing prior work.}

\noindent \textbf{Evaluation.} In the evaluation domain, we scrutinize the operational effectiveness and efficiency of \gnids across diverse datasets. \emph{RQ2: How do the state-of-the-art models perform on established public datasets?} This entails a rigorous reevaluation of models on benchmark datasets, focusing on their reproducibility and replicability. \emph{RQ3: How do these models generalize to new datasets derived from real-world enterprise environments?} This question assesses the adaptability and scalability of these models when confronted with data from a newly curated large-scale enterprise network, comparing their performance on synthetic versus real-world data. This evaluation provides insights into how these models can be optimized for better generalizability and applicability across various scenarios.

\noindent \textbf{Deployment.} In the deployment domain, we address the practical aspects of deploying \gnids in operational settings. \emph{RQ4: How do these models perform from both temporal and spatial perspectives in a production environment?} This question evaluates the scalability and efficiency of the models when processing extensive network traffic data, a critical consideration for enterprise-level deployments. \emph{RQ5: How resilient are these models to adversarial attacks?} This involves stress-testing the models against sophisticated adversarial perturbations to determine their robustness and reliability in real-world attack scenarios. Understanding the robustness and scalability of these models is crucial for ensuring their effective integration into existing security infrastructures.

%\textbf{Deployment} In the deployment domain, we address the practical aspects of deploying \gnids in operational settings. \emph{RQ4: What are the temporal and spatial performance metrics of these models in a production environment?} This question evaluates the scalability and efficiency of the models when processing extensive network traffic data, a critical consideration for enterprise-level deployment. \pj{it is unclear what a "temporal and spatial" metric is. Is the intention to evaluate the long-term operational performance of the G-NIDS? And how can "temporal and spatial metrics" capture the "scalability" of the models?} \emph{RQ5: How resilient are these models to adversarial attacks?} This involves stress-testing the models against sophisticated adversarial perturbations to determine their robustness and reliability in real-world attack scenarios. Understanding the robustness and scalability of these models is crucial for ensuring they can be effectively integrated into existing security infrastructures. \pj{First, RQ5 is a bit far-fetched: just what is an "adversarial attack"? Plus, how likely is this attack to occur in reality? what is crucial is investigating the robustness of these systems against "likely" threats. The threat model of the adversarial attacks should be defined early on to determine whether the envisioned scenario has any relevance}

We conducted a comprehensive R+R study on five representative systems: \anomale~\cite{caville2022anomal}, \vgrnn~\cite{hajiramezanali2019variational}, \pikachu~\cite{paudel2022pikachu}, \euler~\cite{king2023euler}, and \argus~\cite{xu2024understanding}. Our evaluations utilized both the original experimental setups (artifact re-use) and new setups (artifact re-implementation) to critically assess and extend previous findings. Anomal-E leverages edge features and a graph topological structure. VGRNN uses a hierarchical variational model to  capture both topology and node attribute changes in dynamic graphs. \pikachu leverages temporal graph convolutional networks to model dynamic network behaviors. \euler employs variational graph autoencoders for unsupervised anomaly detection. And \argus incorporates edge features via GNN to improve the extraction of network information. We evaluated these systems on well-known public datasets such as LANL~\cite{lanl}, which contains extensive network activity logs, DARPA OpTC~\cite{optc}, designed to simulate real-world operationally critical threats, and CIC-IDS-2017~\cite{sharafaldin2018toward}, which encompasses number of common network attack scenarios. Additionally, we included a newly collected large-scale enterprise dataset, encompassing a diverse range of network activities and attack scenarios, to thoroughly assess the models' generalizability and robustness in real-world environments.

Our experimental results show significant differences in reproducibility. In the implementation domain, we found that key parameters such as the time window for each snapshot, the embedding dimension, and the detection model have a significant impact on the detection efficiency. Optimizing these parameters can improve the performance metrics. In the evaluation domain, we observed that models exhibited more consistent performance on the LANL dataset compared to others, where discrepancies were pronounced due to suboptimal experimental settings and differing preprocessing techniques. Replication experiments showed that fine-tuning the model parameters improved the performance, but there was a significant performance decline on both the CIC-IDS-2017 dataset and our newly collected enterprise dataset, which indicates challenges in terms of generalization and a higher false positive rate. In the deployment domain, we discovered that while \vgrnn and \argus demonstrated the highest space efficiency, processing up to 70K nodes, models like \pikachu struggled with large-scale datasets, encountering memory limitations and significant time consumption. Additionally, \vgrnn, \euler, and\argus are all susceptible to evasion attacks, particularly on the LANL dataset. To enhance the reproducibility and replicability of\gnids, we also provide several recommendations in this paper.

We summarize our contributions as follows.

\begin{itemize}[leftmargin=*]
	\item[--] We collect a new, large-scale dataset from a real-world network, providing a comprehensive and realistic basis for evaluating \gnids. This dataset bridges the gap between academic research and practical application
	\item[--] We develop a robust evaluation framework that integrates artifact reuse and re-implementation to comprehensively assess \gnids across various datasets and configurations, addressing reproducibility and replicability challenges.
	\item[--] We perform an extensive comparative analysis of \gnids across multiple dimensions, such as detection accuracy, scalability, and computational efficiency, utilizing both public datasets and our new enterprise dataset.
 	\item[--] We identify and analyze the impact of adversarial attacks on \gnids, revealing vulnerabilities and proposing effective mitigation strategies.
	%\item[--] We provide actionable recommendations for future \gnids development.
\end{itemize}

%% file: tex/motivation.tex
\section{Motivation}
\label{sec:motivation}
In this section, we present the background of \gnids, the challenges in R+R of \gnids, and the research questions that motivate our study.

\subsection{Background: \gnids}
\begin{table*}
\centering
\caption{Comparison of SOTA \gnids.\eat{ “Supervised” indicates whether manually annotated data was used during the training process of the model to learn from and optimize the model parameters.}}
\label{tab:NIDS}
\scalebox{0.85}{
\begin{threeparttable}
\begin{tabular}{c|cccccccccc} 
\hline
System                                                                   & \begin{tabular}[c]{@{}c@{}}Graph\\Type\end{tabular} & \makecell{Node\\Embedding}                                    & \makecell{Edge\\Embedding}                                    & \begin{tabular}[c]{@{}c@{}}Graph\\Encoder\end{tabular} & \begin{tabular}[c]{@{}c@{}}Temporal \\Encoder\end{tabular} & Streaming & \makecell{Supervised\\ Learning} & Datasets                                                                          & \begin{tabular}[c]{@{}c@{}}Open\\Sourced\end{tabular}  \\ 
\hline
\begin{tabular}[c]{@{}c@{}}NETWALK~\cite{yu2018netwalk}\\(KDD 2018)\end{tabular}              & Static                                              & YES & NO                                   & \makecell{Network Walk\\Autoencoder}                               & NO                                                         & YES & NO      & \begin{tabular}[c]{@{}c@{}}UCI Messages, Digg,\\arXiv hep-th, DBLP\end{tabular}   & YES                                                   \\ 
\hline
\begin{tabular}[c]{@{}c@{}}ARGA~\cite{venturi2023arganids}\\(ACM 2023)\end{tabular}                 & Static                                              & YES & NO & GAE                                                    & NO                                                         & YES  & NO     & \begin{tabular}[c]{@{}c@{}}CTU-13,\\ToN-IoT\end{tabular}                          & NO                                                    \\ 
\hline
\begin{tabular}[c]{@{}c@{}}\textbf{Anomal-E}~\cite{caville2022anomal}\\(KBS 2022)\end{tabular}             & Static                                              & YES & YES & GraphSAGE                                              & NO                                                         & YES  & NO    & \begin{tabular}[c]{@{}c@{}}NF-UNSW-NB15-v2,\\NF-CSE-CIC-IDS2018-v2\end{tabular}  & YES                                                   \\ 
\hline
\begin{tabular}[c]{@{}c@{}}E-GraphSAGE~\cite{lo2022graphsage}\\(NOMS 2022)\end{tabular}         & Static                                              & YES & YES & GraphSAGE                                              & NO                                                         & YES  & YES     & \begin{tabular}[c]{@{}c@{}}ToN-IoT, NF-TON-IoT,\\BoT-IoT, NF-BoT-IoT\end{tabular} & YES                                                   \\ 
\hline
\begin{tabular}[c]{@{}c@{}}Jbeil~\cite{khoury2024jbeil}\\(IEEE S\&P 2024)\end{tabular}          & Dynamic                                             & YES & YES & TGN                                                    & GRU                                                        & YES & NO      & \begin{tabular}[c]{@{}c@{}}LANL,\\Pivoting\end{tabular}                           & YES                                                   \\ 
\hline
\begin{tabular}[c]{@{}c@{}}\textbf{VGRNN}~\cite{hajiramezanali2019variational}\\(NeurIPS 2019)\end{tabular}            & Dynamic                                             & YES & NO                                   & GCN                                                    & GRNN                                                       & YES  & NO     & \begin{tabular}[c]{@{}c@{}}Cora, Citeseer,\\Pubmed\end{tabular}                   & YES                                                   \\ 
\hline
\begin{tabular}[c]{@{}c@{}}\textbf{EULER}~\cite{king2023euler}\\(NDSS 2022)  \end{tabular}          & Dynamic                                             & YES & NO                                   & GNN                                                    & \begin{tabular}[c]{@{}c@{}}GRU /\\LSTM\end{tabular}        & YES   & NO    &  \begin{tabular}[c]{@{}c@{}}LANL,\\OpTC\end{tabular}  & YES                                                   \\ 
\hline
\begin{tabular}[c]{@{}c@{}}\textbf{PIKACHU} \cite{paudel2022pikachu}\\(NOMS 2022) \end{tabular}                                & Dynamic                                             & YES & NO                                   & \makecell{Random Walk \\+ Skip-gram}                                & GRU                                                        & NO & NO    & \begin{tabular}[c]{@{}c@{}}LANL,\\OpTC\end{tabular}                                         & YES                                                   \\ 
\hline
\begin{tabular}[c]{@{}c@{}}\textbf{ARGUS} \cite{xu2024understanding}\\(IEEE S\&P~2024)\end{tabular} & Dynamic                                             & YES & YES & MPNN                                                   & GRU                                                        & YES  & NO     & \begin{tabular}[c]{@{}c@{}}LANL,\\OpTC\end{tabular}                               & YES                                                   \\
\hline
\end{tabular}
        \begin{tablenotes}
			\small
                \item[*] The ``Streaming'' column indicates whether predictions can be performed incrementally on the ingested network traffic data. 
			\item[*] The bolded systems represent those evaluated in this study. We did not evaluate NETWALK due to its extremely slow efficiency, nor ARGA due to the lack of open-source code, nor E-GraphSAGE due to its supervised training approach. In the case of Jbeil, our efforts to reimplement it based on the public source code encountered significant challenges. Despite our attempt to acquire help from the authors, we were unsuccessful.
	\end{tablenotes}
    \end{threeparttable}
}
\vspace*{-2ex}
\end{table*}

In an enterprise, network traffic is typically audited and stored in network logs, recording user behaviors. By analyzing these logs, security analysts can identify adversarial actions. However, with the growing scale of networks, the volume of traffic is surging rapidly. In a medium-sized enterprise, daily data can reach terabytes, posing extreme challenges in identifying attack behaviors. Furthermore, traditional NIDS are typically rule-based, resulting in a high proportion of attack anomalies remaining undetected. To overcome these limitations, \gnids have been developed as a crucial component of enterprise network security.

The detection rules in traditional NIDS are designed by security analysts, which require constant and costly maintenance. In contrast, \gnids autonomously learn behavioral patterns in network traffic by constructing directed graphs from network communications. In these graphs, nodes represent machine,s and edges represent communication flows. They use graph encoders to generate embeddings of nodes and edges based on normal traffic data and often employ temporal encoders to capture the temporal dynamics of the graphs. Graph Neural Networks (GNNs) are typically leveraged for both graph and temporal encoders. By learning normal traffic patterns, \gnids detect any deviations as suspicious behavior. They show strong detection performance, even against unknown attacks like zero-day exploits, and offer greater adaptability than traditional NIDS. Table~\ref{tab:NIDS} summarizes \gnids proposed recently.
%In recent years, various \gnids have been proposed~\cite{yu2018netwalk, venturi2023arganids, caville2022anomal, lo2022graphsage, khoury2024jbeil, hajiramezanali2019variational, king2023euler, paudel2022pikachu, xu2024understanding}, as summarized in Table~\ref{tab:NIDS}. 

\subsection{Challenges in R+R of \gnids}

\gnids have demonstrated outstanding performance on public datasets, showing great potential in cybersecurity defense \cite{ yu2018netwalk, paudel2022pikachu, king2023euler, kipf2016variational, xu2024understanding,  cheng2021step, cai2021structural}. However, the reproducibility of these results remains unexplored, making it difficult to assess their validity and reliability in different contexts, especially new scenarios.

A key challenge is due to the limited public datasets used for evaluation. While public datasets are representative of real-world environments, significant disparities exist compared to actual industry settings. The main differences include the larger scale of network traffic, more complex network structures, and a wider array of network threats faced by enterprises, which are not fully captured in datasets like LANL and OpTC. Additionally, in real-world environments, NIDS must respond to attacks promptly with limited resources, processing massive amounts of data efficiently while maintaining high accuracy. The robustness of models is also crucial, as they must remain accurate and stable under hostile conditions.

To bridge this gap, we leverage a new dataset collected from an anonymous enterprise to evaluate the adaptability of existing \gnids to contemporary cybersecurity challenges. 
%In recent years, diverse \gnids have been proposed \cite{paudel2022pikachu, king2023euler, xu2024understanding, yu2018netwalk, kipf2016variational, cheng2021step, cai2021structural}. Among them, PIKACHU \cite{paudel2022pikachu}, EULER\cite{king2023euler}, and ARGUS \cite{xu2024understanding} are the latest and most representative. 
We have selected five representative systems (\ie the bolded systems as shown in Table~\ref{tab:NIDS}) for evaluation, as they represent recent advancements and have consistently showed high performance in detection accuracy, scalability, and computational efficiency, demonstrating their superiority in all aspects of network intrusion detection. We re-evaluate these models on both public datasets and our own dataset in terms of effectiveness and efficiency.

\subsection{Our Research Questions}

We introduce our research questions (RQs) related to the above-mentioned challenges below and justify their inclusion in our study.

%\pj{We categorize our research questions into three key areas: implementation, evaluation, and deployment. The implementation domain identifies the most impactful settings for optimal performance. The evaluation domain provides insights into how these models can be optimized for better reproducibility and replicability across various scenarios. And the deployment domain shows the scalability and robustness of \gnids models. }

%\pj{ \textbf{Implementation} In the implementation domain, our focus is on the intricacies of \gnids configurations and their impact on system performance. Specifically, }
\noindent \textbf{RQ1: What are the key factors that influence the re-imple\-mentations of \gnids?} 

Besides the factors discussed in the footnote of Table~\ref{tab:NIDS}, which prevent the re-implementation of \gnids, settings such as model hyperparameters can significantly influence the detection performance of runnable \gnids. Although researchers have evaluated their approaches on public datasets and open-sourced the model code, they may overlook the impact of key parameters (\eg thresholds and learning rate) on the model's performance, posing challenges in reproducing and optimizing results. Hence, we aim to rigorously analyze various settings to understand how different configurations and hyperparameters affect the overall effectiveness of the system.

%\pj{\textbf{Evaluation } In the evaluation domain, we examine the operational effectiveness and efficiency of \gnids across diverse datasets.} 
\noindent \textbf{RQ2: How do the state-of-the-art models perform on established public datasets?} 

Despite detailed results in original papers, \gnids face challenges of uncertainty regarding their reproducibility and replicability, impacting real-world applicability. We seek to evaluate these models on three public datasets (\ie the LANL dataset, DARPA OpTC, and CIC-IDS-2017) due to their representativeness in terms of data scale and network attacks. Our strategy comprises two aspects: 1) Reusing the artifacts of the target models and following the same experimental setup (\ie default values that are publicly available) to evaluate \textit{reproducibility}. 2) Leveraging Grid Search \cite{liashchynskyi2019grid} to find the optimal hyperparameters to achieve the best performance, assessing their \textit{replicability}.
%Upon the introduction of new intrusion detection models, authors typically assess the performance of these models on public datasets. Despite providing details of the evaluation process and open source the model code, some evaluation specifics might be overlooked due to space constraints or oversight, such as the settings of certain model hyperparameters and the determination of detection thresholds. While these details may seem trivial, they could significantly impact the detection performance of the model. Additionally, the actual deployment environment of the model may differ from the evaluation environment described in the original paper, potentially leading to discrepancies in detection results.
%In this RQ, our aim is to rigorously reproduce the model evaluation experiments on the LANL and OpTC datasets according to the parameter settings provided in the original papers. We will then compare the reproduced evaluation results with the original evaluation results of the corresponding articles to assess whether there are significant differences and analyze the factors that contribute to any discrepancies. Furthermore, given the ongoing criticism of learning-based detection models due to their low interpretability of alerts and high false positive rates, we will focus on the number of false positives generated by existing detection models on public datasets. We will analyze the potential additional burden posed by these false positives.

\noindent \textbf{RQ3: How do these models generalize to new datasets derived from real-world enterprise environments?} 

Due to the complexity of network communication, datasets collected from different environments could differ significantly. These differences manifest in data size, network structure, and types of attacks, which pose additional challenges for the model's performance. To evaluate the generalizability of the target models, we leverage a new dataset that is collected from industrial arena and simulate new types of network attacks. The statistical information of all datasets utilized in our study is shown in Table~\ref{tab:dataset}.

\begin{table}[!t]
    \centering
    \small
    \caption{Statistics of three public datasets and our newly collected large-scale enterprise dataset.}
    \label{tab:dataset}
    \vspace{-2ex}
    \begin{tabular}{cccc}
    \hline
        Dataset & \# Hosts or IPs & \# Events & Duration (\# days) \\ \hline
        LANL & 17,649 & 1,051,430,459 & 58 \\ 
        DARPA OpTC & 814 & 92,073,717 & 8 \\ 
        CIC-IDS-2017 & 19,129 & 2,830,742 & 5 \\
        Our Dataset & 18,425,098 & 10,338,002,425 & 101 \\
        %1 Day Dataset & 604,754 & 144,160,162 & 1\jp{which dataset?} \\
        \hline
    \end{tabular}
\vspace{-4ex}
\end{table}

%Existing model evaluations are confined to two open-source datasets, which still differ from the actual conditions in industry environments, greatly limiting the model's generalizability. Compared to open-source datasets, the scale of network data in real-world enterprise environments is larger, network structures are more complex, and there is a wider variety of potential network attacks. These differences pose additional challenges for the efficiency and accuracy of intrusion detection models.In this RQ, we will re-evaluate the target model using anonymized data collected from industry environments and simulated various network attacks. We will focus on assessing the differences in detection accuracy of the model across different datasets.

%\pj{\textbf{Deployment} In the deployment domain, we address the practical aspects of deploying \gnids in operational settings.} 
\noindent \textbf{RQ4: How do these models perform from both temporal and spatial perspectives in a production environment?} 

\gnids require the analysis of the structure of network graphs to generate embeddings, a process that consumes considerable computational resources and is highly dependent on the size of the graph. In real-world applications, the scale of network traffic to be analyzed far exceeds existing public datasets. For example, in the enterprise network we have monitored, the traffic collected for just a single day comprises 600,000 nodes, which is \textbf{34} times larger than the scale of nodes in the LANL dataset (Table \ref{tab:dataset}). In such scenarios, it is unclear to what extent the performance of existing models for training and detection is affected, which reflects their adaptability to large-scale datasets. In this research question, we examine how the time and space required by the target models vary with alterations in the size of the traffic flow. We further determine the maximum scale that the target models can manage under limited memory conditions.
%Learning-based intrusion detection models need to analyze the structure of network graphs to generate embeddings, which consume significant computational resources and are highly sensitive to the size of the graph. In real-world settings, the scale of network traffic to be analyzed far exceeds existing public datasets. For example, in the enterprise network traffic we have collected, the network traffic for just one day includes 600,000 nodes, which is \textbf{34} times the scale of LANL nodes. In such scenarios, the extent to which the performance of existing models for training and detection is affected reflects the adaptability of the models to large-scale datasets.In this RQ, We will analyze how the time and space required by the target model for intrusion detection vary with changes in the size of the traffic flow, and attempt to find the maximum scale that the target model can handle under limited memory conditions.

\noindent \textbf{RQ5: How resilient are these models to adversarial attacks?}

Compared to rule-based systems, besides the lower interpretability of alerts, \gnids are more susceptible to adversarial perturbations crafted by attackers. Researchers have demonstrated that Graph Neural Network-based edge detection models can be evaded by adversarial attacks~\cite{lin2020adversarial, zugner2018adversarial, zügner2024adversarial, xu2019topology, chen2018link}. For instance, when \euler is used as the target model and the LANL dataset as the evaluation dataset, injecting only ten access records into the testing dataset can successfully fool the detection system~\cite{xu2023cover}. In the real world, the application prospect of a model is determined by its robustness to adversarial attacks. In this research question, we focus on assessing the robustness of the target models and how their performance is affected under adversarial attacks in both public datasets and the industrial dataset we have collected.

%% file: tex/approach.tex
\section{Approach}
\label{approach}

\begin{figure}[!tbp]
    \centering
    \includegraphics[width=\linewidth]{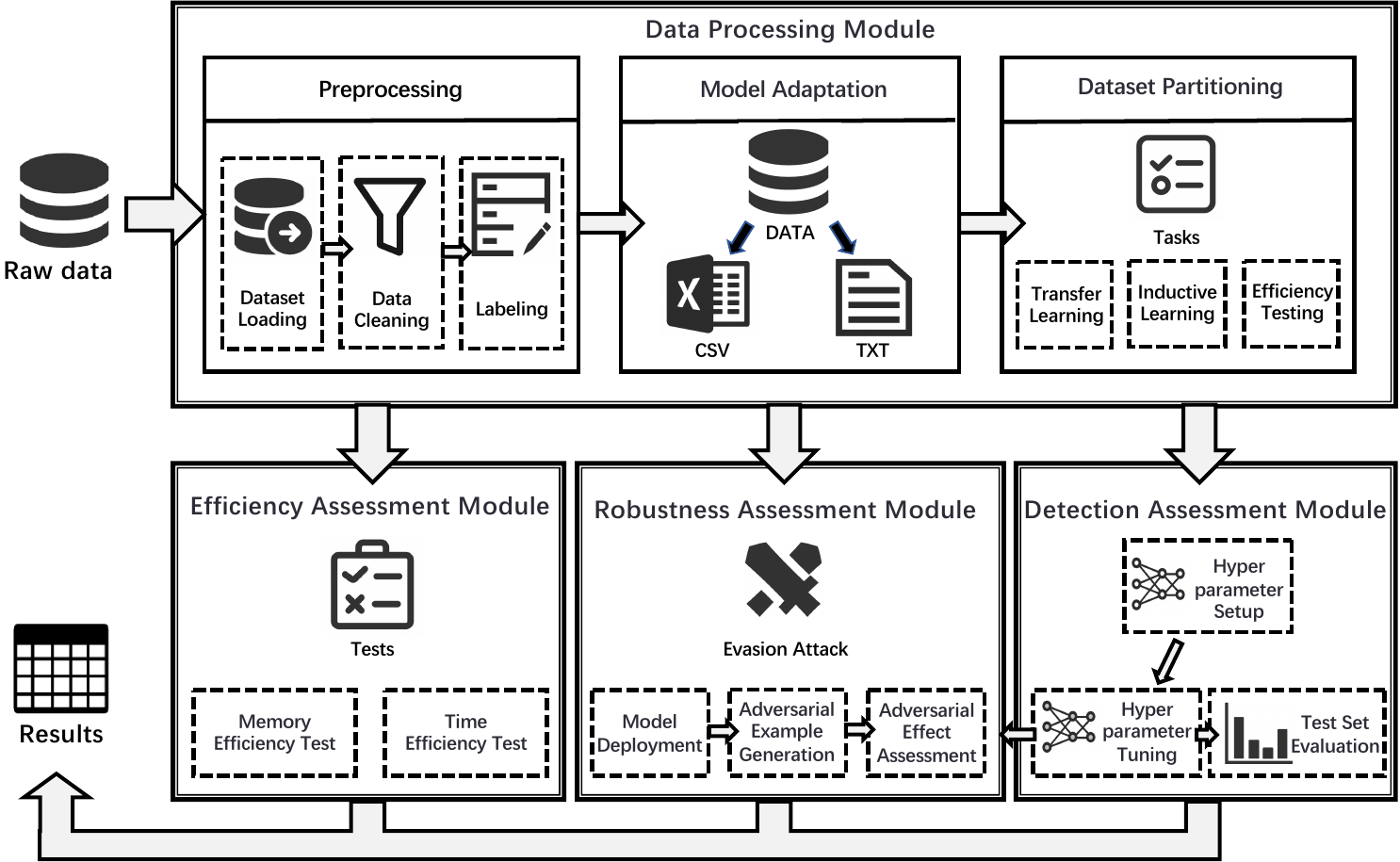}
    \caption{Workflow of \toolname}
    \label{fig:workflow}
\vspace{-4ex}
\end{figure}

We realize our study as a framework, \toolname, whose workflow is shown in Figure~\ref{fig:workflow}. The input to \toolname is the raw data, and the output consists of experimental results. \toolname comprises four key modules. The first module is the Data Processing Module, which preprocesses the data, adapts target models to the data, and divides the data into subsets for efficient processing in subsequent modules. The second module is the Detection Assessment Module, which assesses model performance by reusing or fine-tuning heyperparameters. The third module is the Robustness Assessment Module, which evaluates the robustness of target models under adversarial attacks. The fourth module is the Efficiency Assessment Module, which measures model performance in terms of both space and time. Below we delve into the design details of the modules within \toolname.

\noindent \textbf{Data Processing Module.} The first step in this module involves loading raw data collected from various real-world sources, including public datasets (LANL, OpTC, and CIC-IDS-2017) and a new large-scale enterprise dataset. Real-world data often contains noise or erroneous information, potentially compromising the quality of model training. To ensure clean input for the models, techniques such as filtering invalid entries, eliminating outliers, and smoothing time-related data are employed. Moreover, each event within the dataset is labeled as either normal or malicious. For public datasets, the labels come from detailed red team documentation or the pre-existing labels supplied within the dataset. In the case of the new enterprise dataset, we simulate attacks and record accurate timestamps and related traffic logs, allowing for precise labeling of the data. Additionally, the new enterprise dataset is further divided into smaller subsets to facilitate efficient validation and testing. This helps determine the optimal model parameters without consuming excessive computational resources. Given the variation in datasets from different sources, their formats must be adapted to meet the model's input requirements. For example, models such as \euler and \argus take input in the form of batched text files, whereas models like \pikachu expect CSV files. Finally, the dataset is partitioned based on the specific training tasks. For inductive learning and efficiency testing tasks, the training set consists of all event logs prior to the first attack event timestamp. This allows the model to learn from data prior to any attack occurrences.

\noindent \textbf{Detection Assessment Module.} We first adopt the original experimental setup provided by each model in the original paper, reusing the hyperparameters and dataset configurations to ensure consistency with prior research. This methodology allows us to replicate the original findings and establish a baseline for comparison. Then, we apply grid search techniques to explore a range of hyperparameters in order to identify the optimal configuration on the testing dataset. Our goal is to optimize detection accuracy while minimizing both false positives and false negatives. This step is pivotal for improving the model's performance beyond its original setup. Finally, we evaluate the model's performance using various metrics (see Section~\ref{evaluation}) under the optimal hyperparameter configuration. 
%These metrics include True Positive Rate (TPR), False Positive Rate (FPR), Precision, Average Precision (AP), and the Area Under the Receiver Operating Characteristic Curve (AUC), as delineated in Section~\ref{evaluation}. 
Notice that for the large-scale enterprise dataset, we directly utilize the testing set to ensure the model's optimal performance on this dataset, mirroring its practical performance in enterprise settings.
%It is important to note that our task is to assess whether the model can be applied to real-world enterprise scenarios. Therefore, during the hyperparameter tuning process, we directly use the test set to ensure the model's optimal performance on this dataset, reflecting how it would perform in practical enterprise settings.

\noindent \textbf{Robustness Assessment Module.} Through simulating white-box adversarial attacks, we apply perturbations aimed at evading detection to the testing set. These perturbations are designed to challenge the model's ability to detect threats and test its resilience under adversarial conditions. Utilizing the optimal parameter model obtained from the Detection Assessment Module, we conduct evasion attack testing. This allows us to evaluate the model's performance in detecting attacks, even when subjected to adversarial perturbations. The goal is to assess how effectively the model maintains its detection accuracy under conditions designed to deceive it. This process helps identify potential vulnerabilities within the model and highlights areas that need improvement, particularly with regard to robustness. By understanding the model's weaknesses in the presence of adversarial interference, we can develop strategies for enhancing its defense mechanisms and ensure that it remains reliable and effective in real-world, adversarial environments.

\noindent \textbf{Efficiency Assessment Module.} During the entire training and testing phases, we monitor memory usage closely. By incrementally increasing the number of nodes (hosts) in the dataset, we determine the maximum scale the model can handle before encountering out-of-memory (OOM) errors. This helps identify the limits of the model's scalability and resource requirements. Additionally, we record both the training and testing durations to assess how efficiently the model utilizes computational resources. This timing data provides valuable insight into the computational efficiency of each model. Comparing models based on these metrics highlights the differences in speed and scalability, emphasizing those models that perform well under resource constraints. It also helps assess the feasibility of deploying each model in real-world enterprise environments, where computational resources may be limited. This evaluation ensures that we select models that not only deliver high performance but also operate efficiently in practical settings.

%% file: tex/evaluation.tex
\section{Evaluation Results}
\label{evaluation}
In this section, we discuss the details of the experiments we conducted to address each of the RQs defined in Section~\ref{sec:motivation}. For each RQ, we describe the approach we used to capture the relevant metrics, present the results we obtained, and discuss the implications of these results with respect to each of the RQs. 

\noindent \textbf{Dataset Preparation}:
To evaluate the performance of SOTA \gnids on established public datasets, we selected three well-known public datasets: LANL~\cite{lanl} , DARPA OpTC~\cite{optc} and CIC-IDS-2017~\cite{sharafaldin2018toward}, as shown in Table~\ref{tab:dataset}. The LANL dataset originates from the internal corporate network of the Los Alamos National Laboratory. It spans 58 days and comprises log files from five distinct sources, capturing both regular operational activities and a series of controlled red-team exercises designed to simulate network attacks. The OpTC dataset is a comprehensive collection of network and system logs designed to support research in cybersecurity, particularly in the areas of threat detection and response. This dataset includes logs from about 1,000 machines over one week, capturing both normal operations and simulated network attacks. The CIC-IDS-2017 dataset is a network traffic dataset spanning five days and includes both benign and attack traffic, resembling real-world scenarios.

Furthermore, to assess the generalizability of these \gnids on our newly collected dataset, we first simulated three types of attacks and collected attack traffic. Then, we adopted the same approach as in~\cite{gui2025netguardian} to merge attack traffic into the new dataset. Table~\ref{tab:attacks} in the appendix shows the statistical information of the simulated attacks. We configured the network environment of the victim enterprise in virtual machines and simulated the attacks following the guidelines provided by the MITRE adversary emulation library~\cite{mitretool}. This environment comprises a Windows domain consisting of a Domain Controller (running Windows Server 2019) and multiple domain-joined hosts running Windows and Linux systems. We chose three representative APTs: Sandworm, Wizard Spider, and OilRig, known for their comprehensive attack chains and destructive power. The objective of these attacks was to penetrate the domain environment and gain control over the Domain Controller, thereby gaining mastery over the entire domain environment. During the attacks, we utilized Zeek to capture attack traffic flows as attack events.

It is worth noting that several other network attack datasets exist, such as CIC-IDS-2018~\cite{sharafaldin2018toward}, ToN-IoT~\cite{moustafa2015unsw}, BoT-IoT~\cite{moustafa2015unsw}, and UNSW-NB15~\cite{moustafa2015unsw}. Howevr, we did not select these datasets due to their limitations. Specifically, they either mix attack and normal behaviors based on timestamps or contain over 90 percent attack behaviors, which prevents them from fulfilling the training requirements of models within the target \gnids.

\noindent \textbf{Environment settings.} We conducted our experiments on a workstation equipped with an Intel i9-14900K 32-core processor and 128 GB of CPU memory, running the Ubuntu 22.04.4 LTS operating system. RQ1 to RQ4 were conducted using the CPU, while RQ5 used the GPU. Our GPU was an NVIDIA GeForce RTX 4090 with 24 GB of memory. The runtime environments for the models were consistent with those in the original papers.% \euler used PyTorch 1.13.1 and Python 3.10.8; \argus used PyTorch 1.10 and Python 3.9.12; and \pikachu used TensorFlow 2.2.0 and Python 3.6.15.

\noindent \textbf{Evaluation metrics.} Similar to previous works~\cite{king2023euler, xu2024understanding}, we define the edges in the traffic graph that contain at least one malicious event as true positives (TP) and edges containing all normal events as true negatives (TN). False positives (FP) and false negatives (FN) are defined as the edges that are misclassified as malicious and normal, respectively. Malicious events in the LANL and OpTC datasets are extracted based on the detailed red-team documentation. In our constructed evaluation dataset, simulated attacks on our hosts captured the start and end times of each stage and related traffic, excluding benign traffic during attack execution for accurate labeling. All datasets underwent manual verification for labeled edge completeness.

To determine the performance of target models, we select the following metrics in our evaluation: true positive rate (TPR) that is equivalent to recall, false positive rate (FPR), precision, average precision score (AP), and the area under the ROC curve (AUC) that plots TPR against FPR for different thresholds. The first four metrics are defined as follows.

\vspace{-3ex}
\begin{equation*}
\begin{aligned}
    \text{TPR} &= \frac{TP}{TP + FN} & \hspace{20pt} \text{FPR} &= \frac{FP}{FP + TN} \\
    \text{Precision} &= \frac{TP}{TP + FP} & \hspace{20pt} \text{AP} &= \sum\limits_{\tau} (R_\tau - R_{\tau-1}) P_\tau
\end{aligned}
\vspace{-2ex}
\end{equation*}
, where $R_\tau$ and $P_\tau$ represent the recall and precision scores at the threshold $\tau$, respectively.

%In recent work, Xu \etal~\cite{xu2024understanding} pointed out that due to the highly imbalanced number of attack and normal events in the dataset, the AP score is more suitable for measuring model performance than the AUC score.

%\begin{custombox}{RQ1}
%What key implementation parameters influence the detection efficacy of Graph NIDS?
%\end{custombox}
{\vspace{4pt} \setlength{\parindent}{0pt} \large \textbf{RQ1: What are the key factors that influence the re-implementations of \gnids?}\vspace{3pt}}

\noindent \textbf{Approach:} To address this research question, we leveraged the Detection Assessment Module to fine-tune the implementation parameters of the target models and evaluated them using the OpTC dataset. We focused on the OpTC dataset because it is the common dataset utilized in the original papers of the target models. Tables \ref{tab:para-anomale}-\ref{tab:para-argus} in the appendix show the specific implementation parameters. We adopted the sum of AP and AUC as the criterion for parameter optimization. After obtaining the optimal settings, we assessed the detection efficacy of \gnids by adjusting a specific parameter while maintaining all other settings at their optimal values. Then, we meticulously analyzed the impact of the implementation parameters on the model's performance, and identified those having a significant influence on the overall effectiveness of the target models as the key parameters.

\begin{figure}[!tbp]
    \centering
    \includegraphics[width=\linewidth]{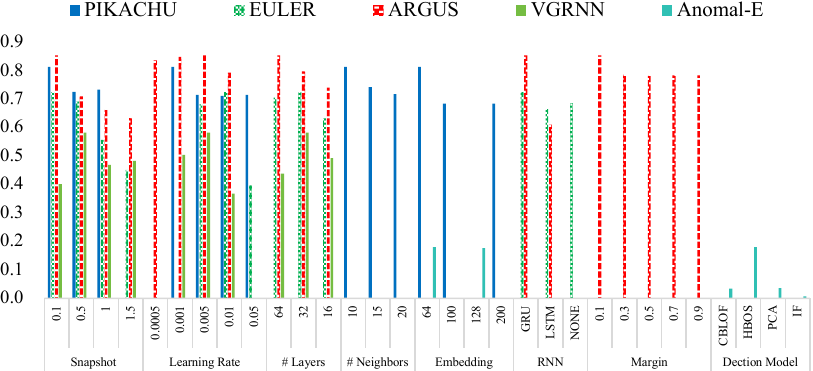}
    \caption{Impact of key implementation parameters on AP. Notice that some models lack results corresponding to certain parameters due to the absence of those parameters. Table~\ref{tab:parameter-description} in the appendix explains each parameter on the X axis.\eat{ ``Snapshot'' denotes the time window of each snapshot. ``\# Layers'' denotes the number of layers in the GNN model. ``\# Neighbors'' denotes the number of sampled neighbors. ``Embedding'' denotes the dimension of embedding.}}
    \label{fig:para}
\vspace{-2ex}
\end{figure}

\noindent \textbf{Results:} 
Figure~\ref{fig:para} shows the results of the impact of key implementation parameters. For ease of comparison, we have also included the results of these parameters (if applicable) across models. We focus on two metrics, AP and AUC, which are utilized for assessing the performance of the models in their original papers. %For \pikachu, we adjusted the model's learning rate and the number of sampled neighbors when generating joint embeddings. For \euler and \argus, we adjusted the time window size of each snapshot, the model's learning rate, and the number of GNN layers.

\noindent \textbf{Discussion:} 
For \vgrnn, \euler, and \argus, the size of the snapshot significantly impacts their performance. Specifically, as the time window of the snapshot increases, the AP scores of \euler and \argus decrease markedly, whereas the AP score of \vgrnn initially rises and then drops significantly. The underlying reason is that when encoding each snapshot using a GNN, all events between two nodes are compressed into a single edge, leading to a loss of temporal dynamics within the snapshot. With a larger snapshot, the edge/event ratio dminishes, making it more difficult for \gnids to distinguish between attack events and normal events as they are compressed together. Conversely, if the snapshot is too small, the contextual information contained within it will be reduced, thus limiting the behavioral features that the model can learn. Therefore, setting an appropriate snapshot size for \gnids is crucial.

In addition, the learning rate has a notable affect on the performance of \euler, but has little impact on the remaining models. For \pikachu, the embedding dimension can influence its detection performance. For \anomale, the choice of the anomaly detection model has the greatest impact on the detection effect. However, we observe that these parameters exert a relatively minor influence on the AUC metric, as shown in Figure~\ref{fig:auc} in the appendix.

\begin{custombox}{Finding:}
For \gnids, the size of the snapshot has a significant impact on their detection performance in terms of the AP score, while other parameters exhibit varying degrees of influence. Nonetheless, these implementation parameters have minimal influence on the AUC metric. This observation aligns with existing works~\cite{hajiramezanali2019variational,king2023euler, king2023euler2, xu2024understanding, paudel2022pikachu} that suggest AP be a primary optimization goal for enhancing the detection efficacy of \gnids.
%These results shed light on performance optimization through various implementation parameters, with average precision as a major optimization goal for the detection efficacy of \gnids~\cite{hajiramezanali2019variational,king2023euler, king2023euler2, xu2024understanding, paudel2022pikachu}.
\end{custombox}

%\begin{custombox}{RQ2}
%How do the state-of-the-art models perform on established public datasets?
%\end{custombox}
{\setlength{\parindent}{0pt} \large \textbf{RQ2: How do the state-of-the-art models perform on established public datasets?}\vspace{3pt}}
\label{eval:public-performance}

\noindent \textbf{Approach}: To validate the evaluation results of existing GIDS, we conducted Reproducibility and Replication (R+R) experiments on established public datasets. %Moreover, we verified the effectiveness of the models using new public datasets(CIC-IDS-2017).

In the reproduction experiments, we adopted the same environmental settings as in the original papers. To do it, we utilized the publicly available source code of the target models and their datasets for evaluation. For the LANL dataset, the target models employed different processing strategies. For example, \euler utilized all events marked with NTLM (Windows New Technology LAN Manager) for 58 days as input. \argus, on the other hand, only took the events marked with NTLM within the first 14 days as input. \pikachu deleted events related to certain types of users (\eg administrators), and sampled the remaining normal user events as input. Since the preprocessed LANL dataset is publicly accessible among all models, we directly re-used these separate datasets for evaluation. However, for the OpTC dataset, the data preprocessing scripts are not provided for all models. After contacting the authors, we only received these scripts from the authors of \argus. In contrast, the authors of \euler and \pikachu provided the raw OpTC dataset, which could not be directly used as input for the models. Therefore, we opted to use the dataset preprocessed by \argus as input for all models.

In the replication experiments, we utilized the same datasets but adjusted the model parameters to optimize performance. we evaluated the detection performance of the models by fine-tuning their parameters. To train and validate \argus and \euler, we adjusted parameters such as the number of GNN layers, snapshot size, number of epochs, learning rate, threshold weight, and patience. During the testing phase, we varied margin parameters to measure the performance of \argus in terms of average precision loss, and we applied different RNN models to assess the performance of \euler. For the training of \pikachu, we adjusted the embedding dimension of nodes, the learning rate, and the number of neighbors sampled. Through these experiments, we obtained the optimal parameters for the target models on public datasets. The specific parameters are listed in Tables~\ref{tab:para-anomale}-\ref{tab:para-argus} in Appendix~\ref{parameters}.

Notice that since the \anomale model requires the utilization of edge features for each communication, and the LANL and OpTC datasets are unable to generate statistically significant edge features for individual logs, the \anomale model was exclusively tested on the CIC-IDS- 2017 dataset.

%Regarding the evaluation metrics, there were differences among the existing works, such as the absence of AP and precision in \pikachu. To better showcase the models' performance, we selected the union of these evaluation metrics.

\begin{figure}[!tbp]
    \centering
    \begin{subfigure}{0.4\textwidth}
        \centering
        \includegraphics[width=\linewidth]{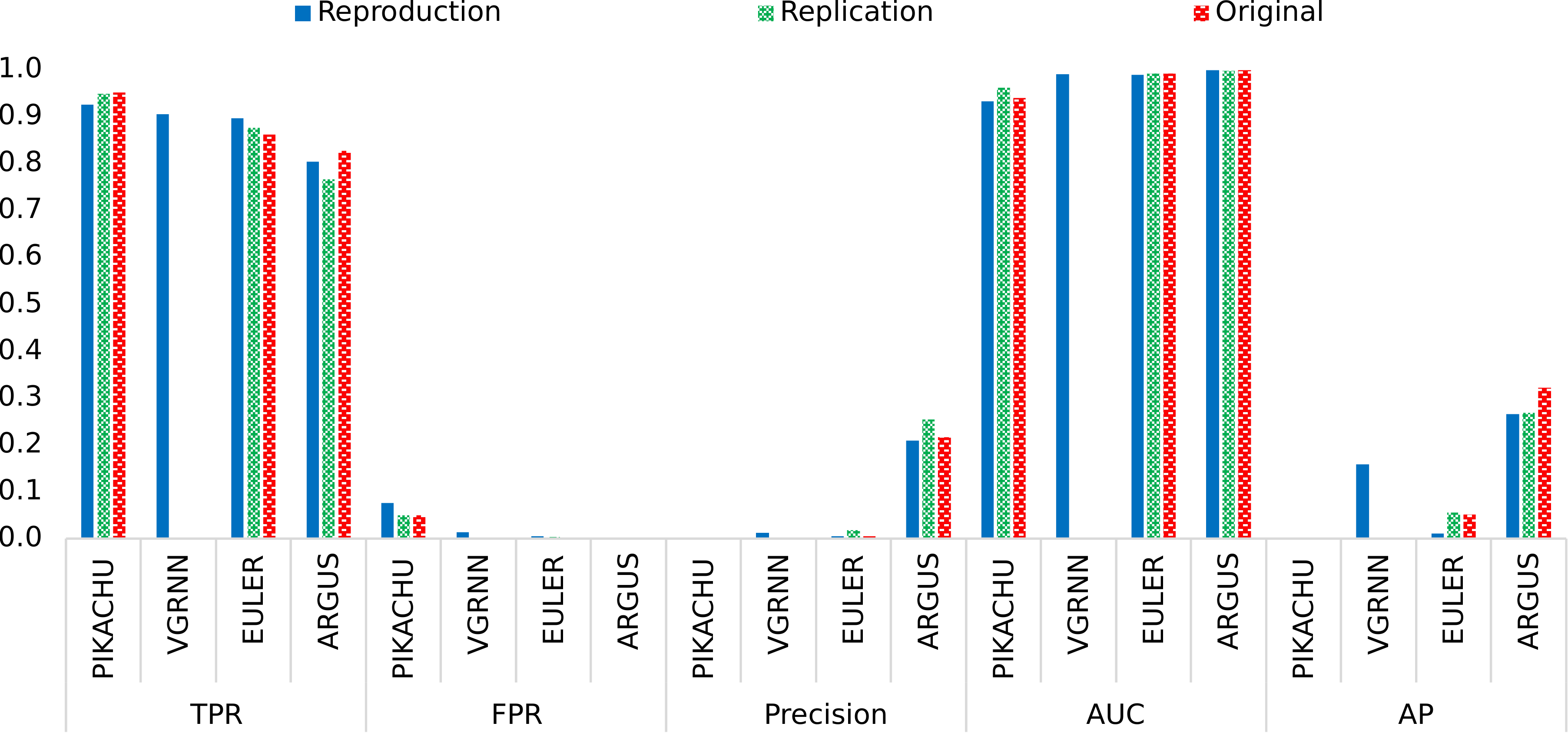}
        \caption{LANL results}
        \label{fig:rq1-lanl}
    \end{subfigure}
    \hfill
    \begin{subfigure}{0.4\textwidth}
        \centering
        \includegraphics[width=\linewidth]{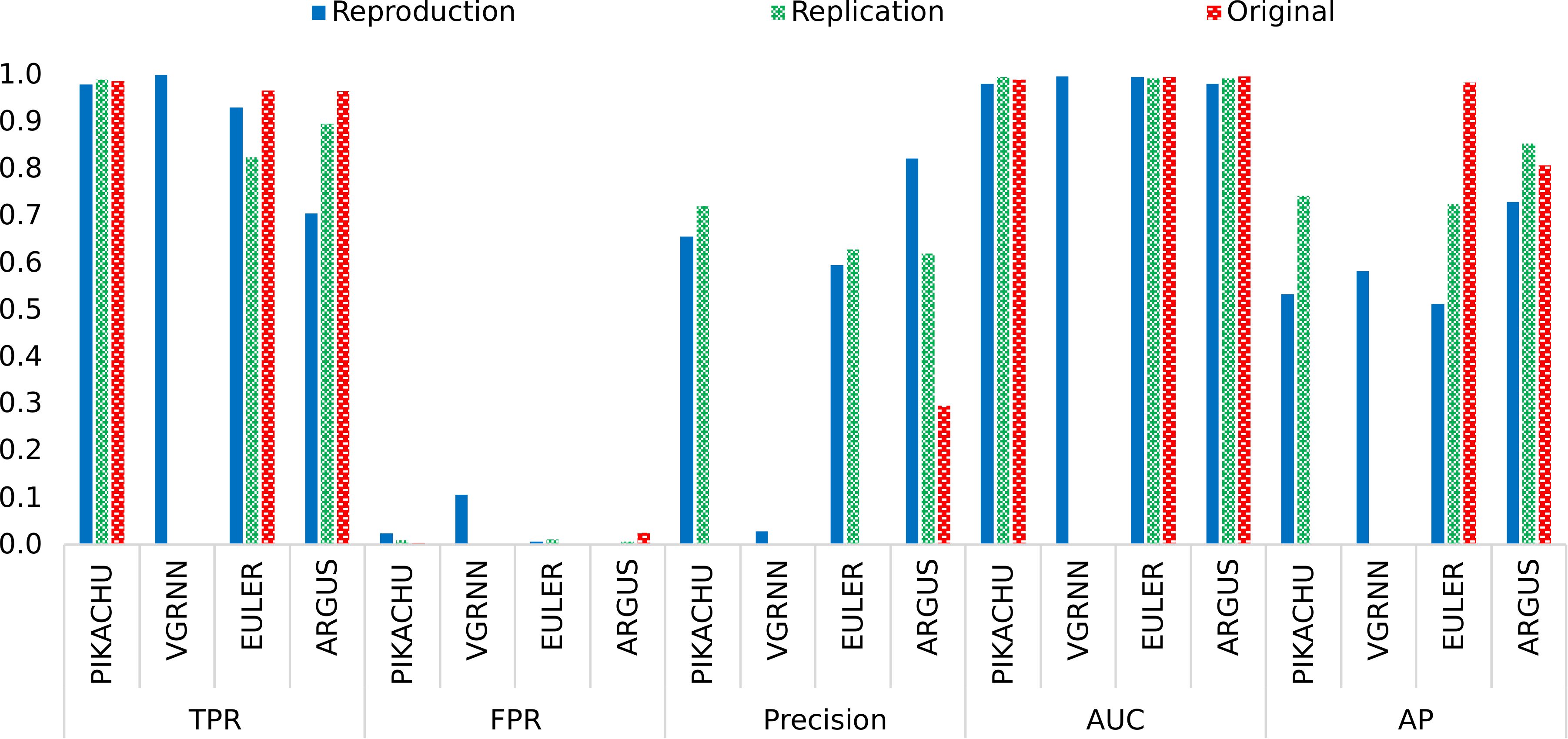}
        \caption{OpTC results}
        \label{fig:rq1-optc}
    \end{subfigure}
    \caption{Performance comparison of target models on the LANL and OpTC datasets. Notice that some original results of \vgrnn, \pikachu and \euler are not plotted in the figure since they are not provided in their original papers.}%Performance comparison of target models on the LANL and OpTC datasets. Notice that we did not plot the original accuracy and AP scores of PIKACHU on the two datasets, nor the original results of VGRNN. Additionally, we did not plot the original accuracy score of EULER on the OpTC dataset because they were not included in their original papers. \cite{hajiramezanali2019variational,paudel2022pikachu, king2023euler2}.
    \label{fig:rq1}
\vspace{-2ex}
\end{figure}

\noindent \textbf{Results:} The results of R+R experiments on the LANL and OpTC datasets are shown in Figure~\ref{fig:rq1}. For ease of comparison, we have aggregated the experimental results and the original ones along the X axis. For each of the metrics, each model has three bars, presented from left to right as the reproduction, replication, and original results. The experimental results on the CIC-IDS-2017 dataset are shown in Table~\ref{tab:rq1} in the appendix.
%Additionally, since the original results of \pikachu do not include AP and precision, and the original results of \euler do not include the precision score on the OpTC dataset, we did not plot them in the figure.

\noindent \textbf{Discussion:}
Overall, the performance of \pikachu,\argus and \euler is more consistent on the LANL dataset than on the OpTC dataset. Among the results, the performance difference in some metrics can be significant, even for the same model. Below we present a detailed analysis of the performance comparison.

In the reproduction experiments, the results of all metrics except AUC on the OpTC dataset showed significant differences with the original results. We identified multiple reasons that could contribute to such differences: 1) The experimental settings that are publicly available are not optimal, which is also reflected by the replication results. 2) The preprocessed dataset of ARGUS which we utilized as input for all models might differ from the ones used by PIKACHU and EULER in their original evaluations. 3) There is randomness in the model's calculation of the detection threshold. In anomaly detection, the model generates a score for each edge based on learned behavioral features, automatically calculates a threshold using the validation set, and classifies edges with scores below this threshold as anomalies. However, EULER and ARGUS randomly select 5\% of events in each snapshot as the validation set. Hence, the calculated threshold can vary significantly, leading to fluctuations in threshold-related evaluation metrics (TPR, FPR and precision). For example, in ARGUS, both TPR and FPR scores are lower than the original results. 

The results also demonstrate that compared to other metrics, AUC is insufficient to assess R+R of \gnids. In addition, we observed different versions of EULER were evaluated in the original paper~\cite{king2023euler2}. We further conducted reproduction experiments of these versions on the OpTC dataset, and found our results are consistent with those in the ARGUS paper, but significantly different from those in the original paper of EULER~\cite{king2023euler2}. We outline more discussions on it in Appendix~\ref{compare}.

In the replication experiments, considering the randomness in calculating detection thresholds, we focused more on the AUC and AP scores, which are independent of threshold calculation. We selected the parameters that yielded the highest AUC and AP scores as the best tuning results. As shown in Figure~\ref{fig:rq1-lanl}, in the LANL dataset, the performance of the target models was relatively stable, and the optimal results from parameter tuning were close to the parameter settings in the original paper. However, in the OpTC dataset, although the results of the replication experiments still differed from the results in the original paper, they were significantly better than the reproduction results. This indicates that evaluating target models in different environments may require re-tuning the parameters.

%In the CIC-IDS-2017 experiment, \pikachu and \anomale performed well, while the remaining models performed poorly. With a high false positive rate, their True Positive Rates (TPRs) were lower than 70\%. This indicates that these models are unable to effectively detect anomalies in the CIC - IDS - 2017 dataset, and they lack versatility in dealing with common network attacks other than Advanced Persistent Threat (APT) attacks. Meanwhile, we will analyze the reasons for the good performance of \pikachu and \anomale together with the experimental results in RQ3. Please refer to Reason 2) in RQ3.

For the experiments conducted on the CIC-IDS-2017 dataset, \pikachu and \anomale exhibited good performance. However, the remaining models performed poorly, with TPRs $\sim$70\% lower than those of \pikachu and \anomale. This indicates that these remaining models lack versatility in dealing with common network attacks included in the CIC-IDS-2017 dataset, beyond the Advanced Persistent Threat (APT) attacks evaluated in their original papers. In RQ3, we further analyze the reasons behind the outstanding performance of \pikachu and \anomale.

\begin{custombox}{Finding:}
SOTA \gnids do not perform consistently across public datasets. Without detailed experimental documentation, such as model parameters, data preprocessing scripts, and environmental settings, it is challenging to accurately reproduce the experiments and validate the results.
%Sharing detailed experimental documentation is crucial, including model parameters, data preprocessing scripts, and environmental settings. This ensures that other researchers can accurately reproduce experiments and validate results.
\end{custombox}

%Additionally, in the experiments on the LANL dataset, we observed that \pikachu first performs random sampling on the original dataset. Specifically, \pikachu retains the 104 anomalous users listed in the redteam file and then randomly samples 2,080 normal users, which is 20 times the number of anomalous users\jp{better to be in the approach part}, using these users' events as the sampled dataset. Subsequently, \pikachu uses the sampled data as input for both the training and testing phases and calculates the false positive rate based only on the number of events in the sampled dataset. Because the sampled dataset contains fewer normal events, this may lead to differences in the model's performance comparison.

%\begin{custombox}{RQ3}
%How do these models generalize to new datasets derived from real-world enterprise environments?
%\end{custombox}
{\setlength{\parindent}{0pt} \large \textbf{RQ3: How do these models generalize to new datasets derived from real-world enterprise environments?}\vspace{3pt}}
\label{eval:commercial-performance}

\noindent \textbf{Approach:} To answer this research question, we constructed an evaluation dataset based on real-world network traffic that is collected from massive networks. Compared to public datasets, this dataset is much larger in scale, encompassing a new scenario with a wider range of behaviors. To prevent exceeding the model's processing limits (see RQ4), we first selected a random day of enterprise network data and employed random sampling to identify a subset of nodes and their corresponding communications as background traffic. The number of chosen nodes is close to that of nodes in the LANL dataset. Then, we created three evaluation datasets, each based on a distinct attack scenario, as input for the target models. We set the attack duration to be in the last four hours of the day and leveraged all snapshots (\ie events within time windows) before the occurrence of the first attack event as the training set, and the remaining snapshots for testing. For \vgrnn, \euler, and \argus, 5\% of the edges in the training set are selected to calculate the anomaly score threshold.

For \pikachu, we followed the method described in the original paper to sample node pairs. Assuming there are $n$ nodes with anomalous communications, we randomly sampled $2,000\times n$ normal nodes. To retain as much communication data related to the anomalous nodes as possible, we first extracted the normal nodes that communicated with the anomalous nodes, denoted as set $V_1$, and the remaining nodes were denoted as set $V_2$. We then sampled 80\% of the nodes from set $V_1$. If the number of sampled normal nodes was still less than $2,000\times n$, we continued sampling from $V_2$. Finally, we extracted all the sampled nodes and their communications as the input for the \pikachu model.

For \argus and \anomale, we also incorporated six communication features to characterize user behaviors for the control experiments conducted in the original paper~\cite{xu2024understanding}. These features include the mean and standard deviation of communication duration, the number of packets, and the number of bytes transmitted.

\noindent \textbf{Results:}
The results of these experiments are shown in Table~\ref{tab:rq2}. Here, ``\argus\_ft'' represents the detection performance of the model when considering communication features between nodes, and ``\argus'' represents the detection performance without accounting for these features. These results are optimized through fine-tuning the target models on three datasets. For ease of analysis, we have bolded the best results for each attack dataset.

\begin{table}
\centering
\caption{Evaluation results on the new dataset.}
    \label{tab:rq2}
    \vspace{-2ex}
    \begin{threeparttable}
    \small
\begin{tabular}{lllllll} 
\hline
\textbf{Attack}               & \textbf{Model} & \textbf{TPR}    & \textbf{FPR}    & \textbf{P}      & \textbf{AUC}    & \textbf{AP}      \\ 
\hline
\multirow{6}{*}{OilRig     }  & ARGUS\_ft      & 0.976~          & 0.176~          & 0.021~          & 0.918~          & 0.022~           \\
                              & ARGUS          & \textbf{1.000}~ & 0.115~          & 0.030~          & 0.942~          & 0.030~           \\
                              & EULER          & \textbf{1.000}~ & 0.070~          & 0.048~          & 0.951~          & 0.032~           \\
                              & VGRNN          & \textbf{1.000}~ & 0.065~          & 0.051~          & 0.951~          & 0.031~           \\
                              & PIKACHU        & \textbf{1.000}~ & \textbf{0.002}~ & \textbf{0.649}~ & \textbf{0.999}~ & \textbf{0.560}~  \\
                             & Anomal-E       & 0.875~          & 0.006~          & 0.264~          & 0.934~          & 0.231~           \\
\hline
\multirow{6}{*}{\begin{tabular}[c]{@{}c@{}}Sand-\\ worm\end{tabular}}     & ARGUS\_ft      & \textbf{1.000}~ & 0.383~          & 0.006~          & 0.872~          & 0.009~           \\
                              & ARGUS          & \textbf{1.000~} & 0.227~          & 0.010~          & 0.833~          & 0.006~           \\
                              & EULER          & \textbf{1.000}~ & 0.264~          & 0.017\textbf{~} & 0.835~          & 0.006~           \\
                              & VGRNN          & 0.973~          & 0.243~          & 0.009~          & 0.813~          & 0.006~           \\
                              & PIKACHU        & 0.934~          & \textbf{0.067~} & \textbf{0.017}~ & \textbf{0.961}~ & \textbf{0.016}~  \\
                              & Anomal-E       & \textbf{1.000}~ & 0.115~          & 0.005~          & 0.942~          & 0.005~           \\ 
\hline
\multirow{6}{*}{\begin{tabular}[c]{@{}c@{}}Wizard-\\ Spider\end{tabular}} & ARGUS\_ft      & 0.737~          & 0.153~          & 0.008~          & 0.800~          & 0.010~           \\
                              & ARGUS          & 0.849~          & 0.072~          & 0.017~          & 0.855~          & 0.011~           \\
                              & EULER          & 0.849~          & 0.106~          & 0.012~          & 0.912~          & 0.019~           \\
                              & VGRNN          & 0.849~          & 0.069~          & \textbf{0.018}~ & 0.854~          & 0.012~           \\
                              & PIKACHU        & \textbf{0.967~} & \textbf{0.033}~ & 0.014~          & \textbf{0.998}~ & \textbf{0.548~}  \\
                              & Anomal-E       & 0.984~          & 0.046~          & 0.008~          & 0.969~          & 0.008~           \\
\hline
\end{tabular}
\begin{tablenotes}
\footnotesize
    \item[*] In the table, P denotes precision.
\end{tablenotes}
\end{threeparttable}
\vspace{-4ex}
\end{table}

\noindent \textbf{Discussion:}
Overall, the results show despite a high TPR, the FPR of all models increases significantly, and the precision score drops noticeably, indicating that many false positives were generated during detection. For example, for the Sandworm attack dataset, the \argus model (without edge features) detected 528 attack events but misclassified 1,529,202 normal events as anomalies, resulting in a false positive count that is 2,896 times the number of detected attack events, which is a substantial cost.

In addition, the detection performance of \argus\_ft (considering edge features) decreases in all OilRig and WizardSpider attack datasets and increases in the Sandworm dataset. This is contrary to the trend outlined in the original paper. We observed that the original paper only evaluated the impact of including edge features on the LANL dataset but failed to introduce edge features in another public dataset, OpTC. Therefore, to verify the generalization of incorporating edge features during the training process, we suggest conducting experiments on more datasets. From another perspective, there may be certain attack behaviors whose edge features are similar to normal behaviors, causing the method of introducing edge features to fail.

Surprisingly, on all attack datasets, the detection performance of \pikachu is significantly better than that of \euler and \argus, which is contrary to the conclusions drawn in their original papers. \anomale also demonstrates outstanding performance. There are three possible reasons for this discrepancy. (1) \pikachu and \anomale do not rely on discrete time graphs during the testing phase and do not merge edges within the same temporal graph. Therefore, a high volume of true negatives contribute to the superior detection results of these two methods. (2) \pikachu adopts a distinct threshold-setting approach during evaluation. As discussed in RQ2, \euler and \argus calculate the detection threshold using a validation set randomly sampled from the training set. However, \pikachu directly utilizes the testing set as the validation set (also known as transductive learning) instead of sampling from the training set. After computing the scores for all edges in the testing set, the optimal detection threshold is determined based on the ground truth. This dynamic threshold calculation method, reliant on the ground truth, grants an additional advantage to \pikachu. (3) \pikachu performs additional sampling on the dataset. Restricted by memory limitations, \pikachu cannot process all data. Hence, prior to data analysis, it samples normal communications according to the ratio of normal to abnormal communications. For example, in the Sandworm attack dataset, although the number of attack events remains unchanged after sampling, the number of normal events only accounts for 48\% of the original count, which also provides an additional advantage.Therefore, we used a smaller dataset that included 8,745 normal nodes and their communications, and removed the pre-sampling stage for \pikachu. The results are presented in Figure \ref{fig:rq2-9000} in the appendix. It is evident that although \pikachu still performs the best, the score gap has significantly narrowed.

\begin{custombox}{Finding:}
Most of SOTA \gnids do not generalize well to our new dataset, which is derived from real-world enterprise environments. The performance of recently proposed \gnids may not necessarily surpass that of previous ones. Therefore, to enhance the generalization ability of \gnids, it is necessary to evaluate them on a wider range of intrusion detection datasets to better represent real-world scenarios.
\end{custombox}

% \eat{
% \begin{figure}[!tbp]
%     \centering
%     \includegraphics[width=0.9\linewidth]{fig/RQ4-Memory-Usage.pdf}
%     \caption{Memory Usage. The red horizontal line represents the maximum memory available.}
%     \label{fig:memory1}
% \end{figure}
% }

\begin{figure*}[ht]
    \centering
    \begin{subfigure}[b]{0.32\textwidth}
    	\centering
    	\includegraphics[width=\textwidth]{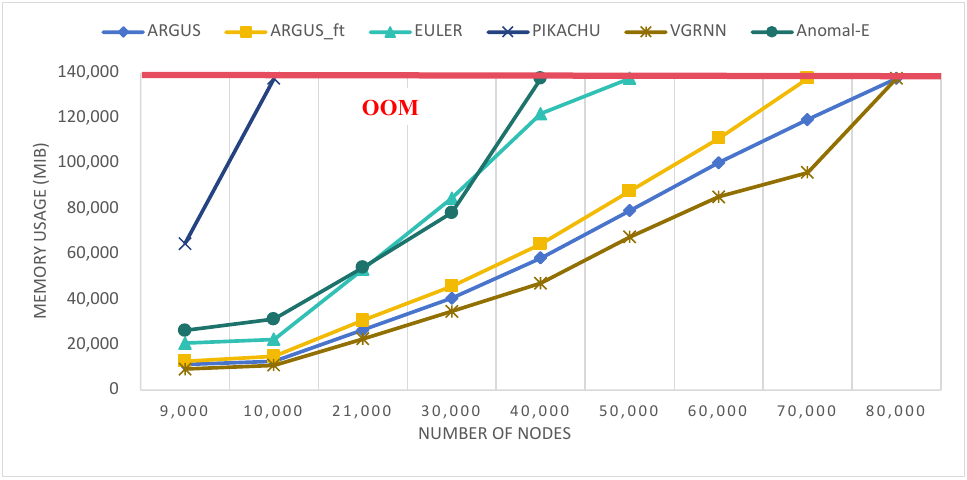}
    	\caption{Memory Usage}
    	\label{fig:memory}
    \end{subfigure}
    \begin{subfigure}[b]{0.32\textwidth}
        \centering
        \includegraphics[width=\textwidth]{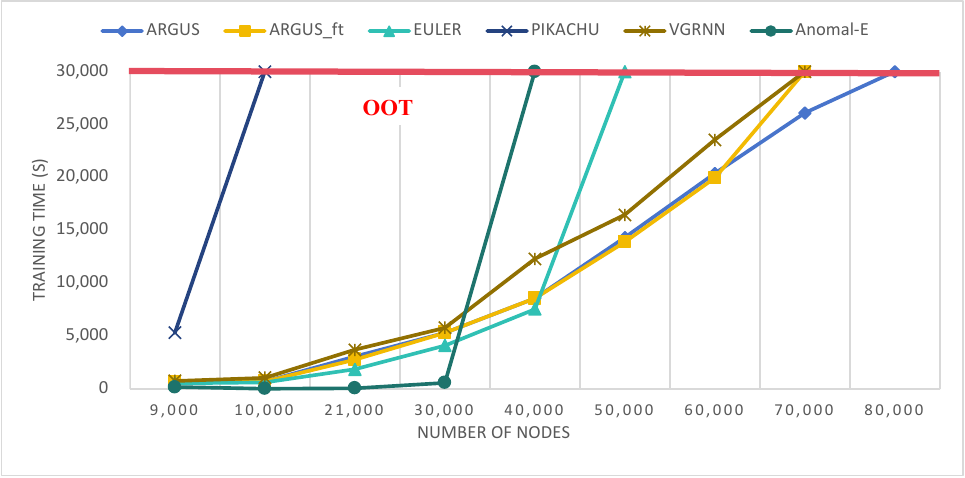}
        \caption{Training Time}
        \label{fig:train-time}
    \end{subfigure}
    %\hfill
    \begin{subfigure}[b]{0.32\textwidth}
        \centering
        \includegraphics[width=\textwidth]{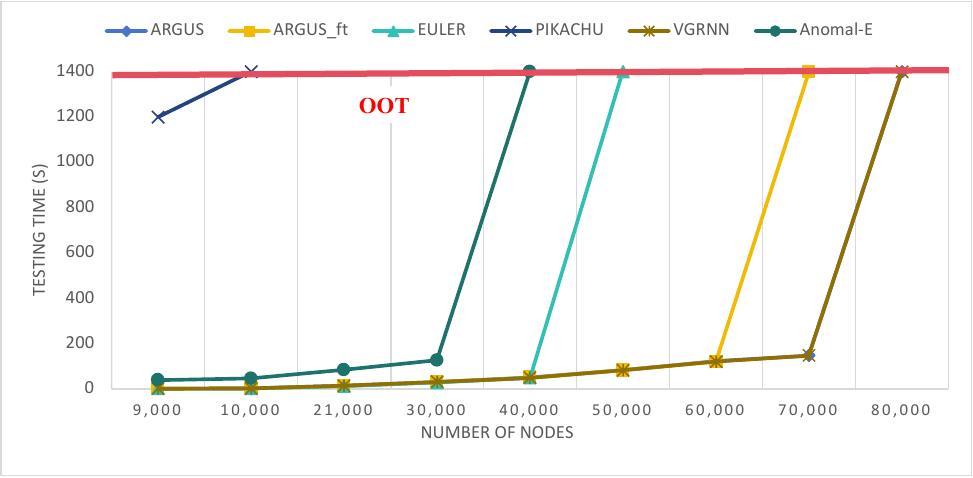}
        \caption{Testing Time}
        \label{fig:test-time}
    \end{subfigure}
    \caption{Memory usage and training/testing time of the target models. The red horizontal line represents the maximum memory or the time budgets available in the experimental environment.}
    \label{fig:cost}
\end{figure*}

%\begin{custombox}{RQ4}
%What are the temporal and spatial performance metrics of these models in a production environment?
%\end{custombox}
{\setlength{\parindent}{0pt} \large \textbf{RQ4: How do these models perform from both temporal and spatial perspectives in a production environment?}\vspace{3pt}}
\label{eval:temporal-spatial-performance}

\noindent \textbf{Approach:} In order to determine the time and space efficiency of the target models, we utilized the Efficiency Assessment Module to measure their running time and memory usage during execution. Specifically, we conducted the evaluation using the network traffic collected from an anonymous enterprise. The enterprise dataset contains voluminous traffc, enabling us to scale the graph size. However, none of the target models can process the entire traffic for a single day in this dataset. Therefore, we randomly chose one day's network data and sampled a subset of nodes along with their communication events.  We gradually increased the number of nodes until it exceeded the processing capacity of the target models. Throughout this process, we recorded the memory usage and time performance of the target models under different dataset sizes, during both the training and detection phases. We partitioned the dataset into training and testing following the same strategy outlined in RQ3.

\noindent \textbf{Results:}
The results are shown in Figure~\ref{fig:cost}, where the X axis represents the number of nodes in the dataset. Figure~\ref{fig:memory} depicts the memory usage of the target models, while Figures~\ref{fig:train-time} and~\ref{fig:test-time} illustrate the training time and testing time, respectively. The red line in the figures represents the maximum memory (128G) or the time budgets (30,000 seconds for training and 1,400 seconds for testing) available in the experimental environment. Reaching this red line indicates that the model will encounter an OOM (Out of Memory) or OOT (Out of Time) error when processing datasets beyond this scale.

\noindent \textbf{Discussion:}
The results show that \vgrnn and \argus have the highest space efficiency. When edge features are disregarded, \argus can manage data with up to 70K nodes. However, incorporating edge features elevates memory overhead, causing OOM errors during the processing of data with 70K nodes. In contrast, \euler can handle up to 40K nodes, exhibiting lower space efficiency compared to \argus. Upon analyzing the design of \euler, we discovered that even when configuring the number of workers and threads to 1, the system still generates both worker and leader processes during runtime. These processes jointly consume memory, and additional memory is utilized for torch inter-process communication, thereby rendering \euler less space-efficient than \argus, which operates serially. \anomale demonstrate low efficiency and can only accommodate 30K nodes. This is mainly because \anomale does not introduce discrete-time graphs and trains the traffic graph as a whole. Conversely, \pikachu's space efficiency is markedly inferior to several other models, encountering OOM errors with more than 10K nodes. This limitation potentially makes it unsuitable for large-scale network data in enterprise environments. The primary reason is that \pikachu processes the entire training set as a single batch during anomaly detection. Large-scale matrix multiplication and differentiation operations lead to a substantial increase in memory usage.

In terms of time efficiency, \argus, \euler, and \vgrnn exhibit similar data processing time during the testing phase. However, \euler requires less time to train datasets of comparable size. We observed that all three models adopt the GNN + RNN architecture. In the RNN component, \argus and \euler utilize the same GRU module, while \vgrnn employs a more intricate GC-LSTM module. Meanwhile, \argus adopts a more sophisticated Message Passing Neural Network (MPNN) for the GNN module, while \euler and \vgrnn utilize a simpler GCN module. Due to the complexity of their model architectures, \argus and \vgrnn experience longer training durations under identical computing resources. Compared with the aforementioned three models, \anomale has longer testing time but shorter training time. This disparity arises because \anomale uses GNN to generate node embeddings, featuring a simpler model architecture. However, during the testing phase, \anomale needs to traverse all edges in the graph simultaneously to generate anomaly scores, which is a time-consuming process. Compared with all other models, \pikachu performs significantly worse in terms of time efficiency. For example, on a dataset with 9,000 nodes, the training time of \pikachu is 9.5 times that of \euler, and the testing time is 537 times longer. This is attributed to \pikachu's two-stage training process: node embedding and anomaly detection training. It first performs random walks on each Dynamic Temporal Graph (DTG) and regenerates node embeddings via GRU. Then, it trains the anomaly detection component based on these embeddings, prolonging the overall training time. During the testing phase, models like \argus and \euler can directly obtain anomaly scores of all edges in a single DTG through reconstruction loss, while \pikachu needs to traverse all edges in the graph to generate adjacent joint embeddings, resulting in substantial time consumption.

\begin{custombox}{Finding:}
\gnids fail to handle large-scale datasets collected from massive networks. The trade-offs between model complexity, memory usage, and computational efficiency highlight the importance of optimizing architecture design for scalability and performance.
%Graph NIDS fail to handle large-scale datasets collected from massive networks. Focus should be on optimizing models for better time and space efficiency on large-scale datasets, refining graph encoding techniques, and exploring more efficient computational algorithms \eg~\cite{yang2022gnnlab, huang2021understanding, jia2020improving, jangda2021accelerating}.
\end{custombox}

%\begin{custombox}{RQ5}
%How resilient are these models to adversarial attacks?
%\end{custombox}
{\setlength{\parindent}{0pt} \large \textbf{RQ5: How resilient are these models to adversarial attacks?}\vspace{3pt}}
\label{eval:robustness}

\noindent \textbf{Approach:} To address this research question, we employed the SOTA evasion attack method~\cite{xu2023cover} targeting GNN-based models in the Robustness Assessment Module. This evasion attack does not modify the training set, but only adds adversarial perturbations to the testing set. In contrast to the poisoning attacks, which necessitate modifications to the training set, evasion attacks are more viable in real-world scenarios as they only require the introduction of additional perturbative behavior during the attack process, without altering the internal network data used for model training. We excluded the evaluations of \pikachu and \anomale for two reasons. Firstly, both the node embedding and anomaly detection processes in \pikachu and \anomale are divided into two distinct stages, and evasion attacks are unable to update the node embeddings. Secondly, and most importantly, the anomaly scores obtained through \pikachu and \anomale are not differentiable with respect to the adjacency matrix. Additionally, we excluded the CIC-IDS-2017 dataset because methods such as \argus are ineffective in detecting anormalies (as discussed in RQ2). Therefore, we applied evasion attacks to three datasets (LANL, OPTC, and our own dataset) using models that had already been trained with optimal parameters. We varied the number of adversarial edges (e.g., 2, 5, 10, 20, and 50), and measured the performance of the target models. To the best of our knowledge, no such attempts have been made in related works.

\noindent \textbf{Results:}
The results for the \vgrnn, \euler and \argus are shown in Table \ref{tab:rq5-vgrnn}, Table~\ref{tab:rq5-euler} (in the appendix) and Table~\ref{tab:rq5-argus} (in the appendix), respectively. In these tables, K represents the number of added adversarial edges. $r_{tgt}$ and $r_{cov}$ denote the average evasion rate of the target event and the covering event, respectively. And $r_{atk}$ denotes the average success rate of the attack. For each target event, the attack is considered successful only if both the target event and the covering event are classified as normal.

\begin{table}[!ht]
\centering
    \caption{Evasion attack performance of VGRNN.}
    \label{tab:rq5-vgrnn}
    \vspace{-2ex}
    \begin{tabular}{c|ccc|ccc|ccc}
    \hline
    \multirow{2}{*}{K} & \multicolumn{3}{c}{LANL} & \multicolumn{3}{c}{OpTC} & \multicolumn{3}{c}{Our Dataset} \\
    & $r_{tgt}$ & $r_{cov}$ & $r_{atk}$ & $r_{tgt}$ & $r_{cov}$ & $r_{atk}$ & $r_{tgt}$ & $r_{cov}$ & $r_{atk}$ \\ \hline
0 & 0.12 & 1.00 & 0.12 & 0.00 & 1.00 & 0.00 & 0.03 & 1.00 & 0.03 \\
2 & 0.68 & 1.00 & 0.68 & 0.00 & 1.00 & 0.00 & 0.04 & 1.00 & 0.04 \\
5 & 1.00 & 1.00 & 1.00 & 0.24 & 1.00 & 0.24 & 0.04 & 1.00 & 0.04 \\
10 & 1.00 & 1.00 & 1.00 & 0.28 & 1.00 & 0.28 & 0.04 & 1.00 & 0.04 \\
20 & 1.00 & 1.00 & 1.00 & 0.75 & 1.00 & 0.75 & 0.04 & 1.00 & 0.04 \\
50 & 1.00 & 1.00 & 1.00 & 0.93 & 1.00 & 0.88 & 0.04 & 1.00 & 0.04 \\ \hline
\end{tabular}
\end{table}

\noindent \textbf{Discussion:}
The results show that \vgrnn, \euler and \argus perform similarly on the LANL dataset. Evasion attacks achieve the highest success rate on the LANL dataset, with all attack events fully covered by the insertion of just two adversarial edges. This underscores the current models' lack of robustness, allowing attackers to evade detection through adversarial attacks. For the OpTC dataset, \euler and \argus nearly cover all attack events by inserting ten adversarial edges.

In contrast, on our own dataset and for \vgrnn on the OPTC dataset, the success rate of adversarial attacks is not high. Even with the insertion of 50 adversarial edges, the attack events cannot be fully covered. A potential reason for this is related to the detection capability of \gnids. As discussed in RQ2 and RQ3, although these target models can detect most of attack events in the dataset, they also exhibit a high false positive rate. This indicates that these models struggle to accurately distinguish between attack events and normal events, resulting in a lowered threshold for anomaly detection. In the adversarial attack experiments, we discovered that the most effective adversarial edges for covering attack events had scores below the detection threshold and were themselves classified as attacks. In contrast, adversarial edges with scores above the detection threshold could not effectively conceal the attack events. From this perspective, a decrease in the model's detection accuracy could paradoxically improve its robustness against adversarial attacks.

\begin{custombox}{Finding:}
\gnids are vulnerable to adversarial attacks. We suggest incorporating defense techniques, such as adversarial detection techniques and robust optimization algorithms~\cite{abusnaina2021adversarial}, to strengthen models’ stability under hostile conditions~\cite{hinton2015distilling, addepalli2020towards, madry2017towards}. 
%GNN-based Graph NIDS are vulnerable to adversarial attacks. We suggest incorporating defense techniques to strengthen models’ stability under hostile conditions, \eg~\cite{hinton2015distilling, addepalli2020towards, madry2017towards, wu2019defending}.  To strengthen Graph NIDS resilience, incorporating adversarial training techniques, where models are trained on both clean and adversarially perturbed data, can improve robustness. Implementing advanced mechanisms, such as adversarial detection techniques and robust optimization algorithms~\cite{abusnaina2021adversarial}, can also mitigate the impact of adversarial attacks in Graph NIDS.
\end{custombox}

%% file: tex/relatedwork.tex
\section{Related Work}

\noindent \textbf{Performance of \gnids.} Various studies have been carried out to analyze the performance of existing \gnids. Most of these studies~\cite{ahmad2021network, bilot2023graph, zhong2024survey} merely reviewed existing evaluation results as analytical basis but did not reproduce or re-evaluate existing NIDS.
%Bilot \etal~\cite{bilot2023graph} reviewed GNN-based NIDS, categorizing NIDS into three types: flow-based, packet-based and authentication-based, and compared the performance of these three types of NIDS. Ahmad \etal~\cite{ahmad2021network} categorized learning-based NIDS into machine learning-based and deep learning-based (\ie graph-based). According to their comparison, the former are more suitable for scenarios with smaller data volumes, while the latter are more suitable for scenarios with larger data volumes. Zhong \etal~\cite{zhong2024survey} reviewed current GNN-based NIDS, discussing the challenges encountered in the processes of graph construction, network design, and GNN model deployment. However, these studies merely reviewed existing evaluation results as analytical basis but did not reproduce or re-evaluate existing NIDS. 
The work by Apruzzese \etal~\cite{apruzzese2023sok} is the most similar to ours. They also conducted extensive evaluation experiments to reveal the gap between research and practice in the NIDS field. However, their work only considered small-scale public datasets and only evaluated traditional machine learning algorithms, such as random forest and logistic regression. In contrast, we assess SOTA \gnids using large-scale datasets, of which three are publicly available and one is commercially collected from real-world production environments. We believe our study complements related works to provide a holistic view of existing \gnids.
%%Overall, there is no related work that reproduces the evaluation results of the latest GNN-based NIDS, while our work not only includes reproduction and replication of existing work but also evaluates the performance of models in real-world production environments through the custom datasets. At the same time, many studies \cite{ahmad2021network, apruzzese2023sok} point out that the evaluation schemes in the current NID field are still incomplete, and the lack of public network datasets containing a sufficient number of complex attacks is a major challenge. We believe that this can be alleviated through the dataset and production methods proposed in this paper.

\noindent \textbf{Robustness of \gnids.} Several studies~\cite{bilot2023graph, aiken2019investigating} have evaluated the robustness of \gnids. Pujol-Perich \etal~\cite{pujol2022unveiling} evaluated the robustness of GNN-based NIDS under two adversarial attacks by modifying packet sizes and arrival times. Their research indicated that learning the relationships between different flows can strengthen the model's robustness. Zhou \etal~\cite{zhou2021hierarchical} proposed an adversarial attack method that significantly reduces the detection accuracy of \gnids in IoT environments. 
%Gardiner \etal~\cite{gardiner2016security} evaluated the machine learning techniques for detecting zombie networks in adversarial settings. 
Apruzzese \etal~\cite{apruzzese2022modeling} analyzed the threat model of existing adversarial attack methods, re-modeled the capabilities of attackers in real-world scenarios, and evaluated the impact of adversarial attacks on NIDS under this threat model. However, most of these evaluation works focused on adversarial attacks established in IoT and SDN environments and only added disturbances at the packet level. In contrast, we leverage adversarial attack methods~\cite{xu2023cover} by incorporating access behaviors into the network to assess the robustness of \gnids.

%% file: tex/conclusion.tex
\section{Conclusion}
Due to the promise of detecting unknown attack patterns such as zero-day exploits, \gnids emerge to provide a modern solution for enterprise security. However, the reproducibility and replicability of these \gnids remain largely unexplored. In this paper, we bridge this gap by systematically evaluating SOTA \gnids on three public datasets\eat{ (LANL, DARPA OpTC and CIC-IDS-2017)} and a newly collected large-scale enterprise dataset. Our findings reveal significant performance discrepancies, highlighting challenges related to dataset scale, model inputs, implementation settings, and robustness against adversarial attacks. Our work provides valuable insights and recommendations for future research, emphasizing the importance of rigorous reproduction and replication studies in developing robust and generalizable \gnids solutions.

%% file: tex/appendix.tex
\appendix
$$\textbf{\LARGE APPENDIX}$$

\begin{table}[!h]
\centering
\caption{Performance comparison of target models on the CIC-IDS-2017 dataset.}
    \label{tab:rq1}
    \vspace{-2ex}
    \begin{threeparttable}
    \small
\begin{tabular}{lllllll}
\hline
\textbf{Dataset}                        & \textbf{Model}     & \textbf{TPR}             & \textbf{FPR}             & \textbf{P}       & \textbf{AUC}             & \textbf{AP}               \\ 
\hline
\multirow{6}{*}{\begin{tabular}[c]{@{}c@{}}CIC-\\IDS-\\2017\end{tabular}} & ARGUS     & 0.600~          & 0.348~          & 0.001~          & 0.702~          & 0.004~           \\
                              & ARGUS\_ft & 0.682~          & 0.177~          & 0.003~          & 0.831~          & 0.004~           \\
                              & EULER     & 0.636~          & 0.254~          & 0.002~          & 0.757~          & 0.006~           \\
                              & VGRNN     & 0.609~          & 0.420~          & 0.001~          & 0.641~          & 0.002~           \\
                              & PIKACHU   & \textbf{0.979~} & \textbf{0.026}~ & \textbf{0.923~} & \textbf{0.977~} & \textbf{0.872}~  \\
                              & Anomal-E  & 0.996~          & 0.231~          & 0.584~          & 0.883~          & 0.583~           \\ 
    \hline  
\end{tabular}
\begin{tablenotes}
\footnotesize
    \item[*] In the table, P denotes precision.
\end{tablenotes}
\end{threeparttable}
\end{table}

\section{Simulated Attacks}
\label{sec:simulated-attacks}
Table~\ref{tab:attacks} presents the statistics of simulated attacks in our own dataset.

\begin{table}[!t]
	\centering
 \small
	\caption{Statistics of simulated attacks.}
	\label{tab:attacks}
        \vspace{-2ex}
	\begin{tabular}{ccc}
		\hline
		Attack Name  & \# Hosts or IPs & \# Events \\ \hline
		OilRig       & 5       & 2532     \\
		Sandworm     & 5       & 587      \\
		Wizard Spider & 4       & 366     \\
		\hline
	\end{tabular}
%\vspace{-2ex}
\end{table}

\begin{table}[t!]
	\scriptsize
	\centering
    \caption{Parameter Description}
	\label{tab:parameter-description}
	\renewcommand{\arraystretch}{1.2}
	\begin{tabular}{@{}c c l@{}}
		\toprule
		\textbf{Parameter} & \textbf{\gnids} & \textbf{Description} \\ 
		\midrule
		Snapshot & \argus, \euler, \vgrnn, \pikachu & Time window of each snapshot.  \\ 
		Learning Rate & \argus, \euler, \vgrnn, \pikachu & Learning Rate.  \\ 
		\# Layers &\argus, \euler, \vgrnn & Number of layers in the GNN model. \\ 
		\# Neighbors & \pikachu & Number of sampled neighbors. \\ 
		Embedding & \pikachu, \anomale & Dimension of embedding. \\ 
		RNN & \argus, \euler & Types of RNN Models.  \\ 
		Margin &\argus &\begin{tabular}[c]{@{}l@{}} Margin parameter used in  \\calculating the average precision loss.\end{tabular} \\ 
		Detection Models & \anomale & Models used for anomaly detection. \\
		\bottomrule
	\end{tabular}
	
\end{table}

\section{Model Parameters}
\label{parameters}
In this section, we will provide a detailed description of the model parameters used in the evaluation experiments. In the replication experiments of RQ2, we adjusted the model parameters on the public datasets LANL , OpTC and CIC-IDS-2017 to achieve optimal results. In the experiments of RQ3, we conducted parameter adjustments on the three selected attack datasets. Specifically, for \anomale, we adjusted the embedding dimension and dection model, with the results shown in Table \ref{tab:para-anomale}. For \vgrnn, we adjusted the number of layers in the GNN model, the time-window size of snapshot, learning rate, threshold weight, and patience, with the results shown in Table \ref{tab:para-vgrnn}. For \pikachu, we adjusted the embedding dimension, snapshot time window size, learning rate, and the number of sampled neighbors, with the results shown in Table \ref{tab:para-pikachu}. For \euler, we tried various combinations of GNN and RNN, and adjusted the number of layers in the GNN model, the time-window size of snapshot, learning rate, threshold weight, and patience, with the results shown in Table \ref{tab:para-euler}. For \argus,  we tried various combinations of RNN, and adjusted the number of layers in the GNN model, the time-window size of snapshot, learning rate, threshold weight, and patience. Additionally, we adjusted the margin parameter used in calculating the average precision loss, with the results shown in Table \ref{tab:para-argus}.

\begin{table}[!tbp]
\centering
\caption{Parameters of \anomale.}
    \label{tab:para-anomale}
    \begin{threeparttable}
\begin{tabular}{lcc}
\hline
Dataset      & Embedding Dimension & Detection Model  \\ \hline
CIC-IDS-2017 & 128                  & CBLOF          \\ \hline
OilRig       & 128                  & CBLOF          \\  \hline
Sandworm     & 64                   & HBOS           \\ \hline
WizardSpider & 64                   & CBLOF          \\ \hline
\end{tabular}
\end{threeparttable}
\end{table}

\begin{table*}[!tbp]
\centering
\caption{Parameters of \vgrnn.}
    \label{tab:para-vgrnn}
    \begin{threeparttable}
\centering
\begin{tabular}{cccccc}
\hline
Dataset      & \# Layers & \makecell{Snapshot \\ Duration (s)} & Learning Rate & Threshold Weight & Patience   \\ \hline
LANL         & 32        & 5400                  & 0.01          & 0.5              & 10         \\ \hline
OpTC         & 32        & 1800                  & 0.005         & 0.5              & 10         \\ \hline
CIC-IDS-2017 & 32        & 600                   & 0.01          & 0.43             & 10         \\ \hline
OilRig       & 32        & 150                   & 0.01          & 0.48             & 10         \\ \hline
Sandworm     & 32        & 150                   & 0.001         & 0.43             & 10         \\ \hline
WizardSpider & 32        & 150                   & 0.005         & 0.48             & 10        \\ \hline
\end{tabular}
\begin{tablenotes}
\footnotesize
    \item[*] In the table, ``\# Layers'' denotes the number of GNN embedding layers
\end{tablenotes}
\end{threeparttable}
\end{table*}

\begin{table*}[!tbp]
\centering
\caption{Parameters of \pikachu.}
    \label{tab:para-pikachu}
    \begin{threeparttable}
\begin{tabular}{cccccc} 
\hline
Dataset      & Embedding Dimension~ &  \makecell{Snapshot \\ Duration (s)} & Learning Rate & Neighbor Number   \\ 
\hline
LANL         & 200                  & 3600                  & 0.001         & 10                \\ 
\hline
OpTC         & 64                   & 360                   & 0.001         & 10               \\ 
\hline
CIC-IDS-2017 & 100                  & 1200                  & 0.001         & 15              \\ 
\hline
OilRig       & 100                  & 300                   & 0.001         & 10               \\ 
\hline
Sandworm     & 100                  & 300                   & 0.001         & 10              \\ 
\hline
WizardSpider & 100                  & 300                   & 0.001         & 10               \\
\hline
\end{tabular}
\end{threeparttable}
\end{table*}

\begin{table*}[!tbp]
\centering
\caption{Parameters of \euler.}
    \label{tab:para-euler}
    \begin{threeparttable}
\begin{tabular}{cccccccc} 
\hline
Dataset      & GNN Model & RNN Model & \# Layers &  \makecell{Snapshot \\ Duration (s)} & Learning Rate & Threshold Weight & Patience  \\ 
\hline
LANL         & GCN       & None      & 64        & 10800               & 0.0005        & 0.6              & 10        \\ 
\hline
OpTC         & GCN       & GRU       & 32        & 360                   & 0.01          & 0.6              & 10        \\ 
\hline
CIC-IDS-2017 & GCN       & GRU       & 32        & 600                   & 0.01          & 0.48             & 10        \\ 
\hline
OilRig       & GCN       & GRU       & 32        & 150                   & 0.005         & 0.48             & 10        \\ 
\hline
Sandworm     & GCN       & GRU       & 32        & 150                   & 0.005         & 0.42             & 10        \\ 
\hline
WizardSpider & GCN       & GRU       & 32        & 150                   & 0.005         & 0.47             & 10        \\
\hline
\end{tabular}
\begin{tablenotes}
\footnotesize
    \item[*] In the table, ``\# Layers'' denotes the number of GNN embedding layers
\end{tablenotes}
\end{threeparttable}
\end{table*}

\begin{table*}[!tbp]
    \centering
    \caption{Parameters of \argus.}
    \label{tab:para-argus}
    \begin{threeparttable}
\begin{tabular}{ccccccccc}
\hline
Dataset                       & Edge Feature & RNN Model & \# Layers & \begin{tabular}[c]{@{}l@{}}Snapshot \\Duration (s)\end{tabular} & Learning Rate & Threshold Weight & Patience & Margin Parameter  \\ \hline
LANL                          & Yes          & GRU       & 32        & 3600                                                            & 0.005         & 0.6              & 10       & 0.8               \\ \hline
OpTC                          & No           & GRU       & 64        & 360                                                             & 0.005         & 0.55             & 10       & 0.1               \\ \hline
\multirow{2}{*}{CIC-IDS-2017} & Yes          & GRU       & 32        & 600                                                             & 0.05          & 0.45             & 3        & 0.8               \\
                              & No           & GRU       & 32        & 900                                                             & 0.01          & 0.46             & 3        & 0.8               \\ \hline
\multirow{2}{*}{OilRig}       & Yes          & GRU       & 32        & 150                                                             & 0.0001        & 0.46             & 3        & 0.8               \\
                              & No           & GRU       & 16        & 450                                                             & 0.01          & 0.48             & 5        & 0.8               \\ \hline
\multirow{2}{*}{Sandworm}     & Yes          & GRU       & 32        & 450                                                             & 0.0001        & 0.45             & 3        & 0.8               \\
                              & No           & GRU       & 16        & 150                                                             & 0.0001        & 0.43             & 3        & 0.8               \\ \hline
\multirow{2}{*}{WizardSpider} & Yes          & GRU       & 32        & 150                                                             & 0.0001        & 0.46             & 5        & 0.8               \\
                              & No           & GRU       & 32        & 150                                                             & 0.001         & 0.48             & 3        & 0.8     \\ \hline         
\end{tabular}
    \begin{tablenotes}
\footnotesize
    \item[*] In the table, ``\# Layers'' denotes the number of GNN embedding layers
\end{tablenotes}
\end{threeparttable}
\end{table*}
\section{AUC with different parameters}
\label{auc}

\begin{figure}[!tbp]
    \centering
    \includegraphics[width=\linewidth]{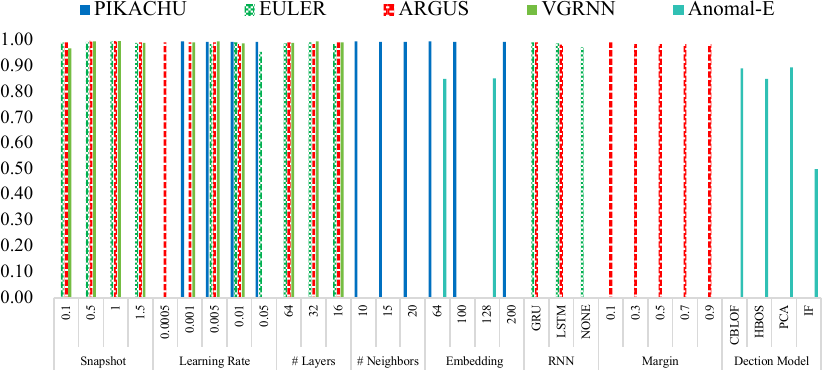}
    \caption{Impact of key implementation parameters on AUC.\eat{ ``Snapshot'' denotes the time window of each snapshot. ``\# Layers'' denotes the number of layers in the GNN model. ``\# Neighbors'' denotes the number of sampled neighbors.}}
    \label{fig:auc}
\end{figure}

The impact of the implementation parameters on AUC scores is shown in Figure~\ref{fig:auc}. It can be seen that with different parameter settings, the models' AUC scores remain stable. Table~\ref{tab:parameter-description} explains each of the parameters on the X axis in Figures~\ref{fig:para} and~\ref{fig:auc}.

\section{Comparison of \euler's Reproduction Results}
\label{compare}
\begin{figure}
    \centering
    \includegraphics[width=0.98\linewidth]{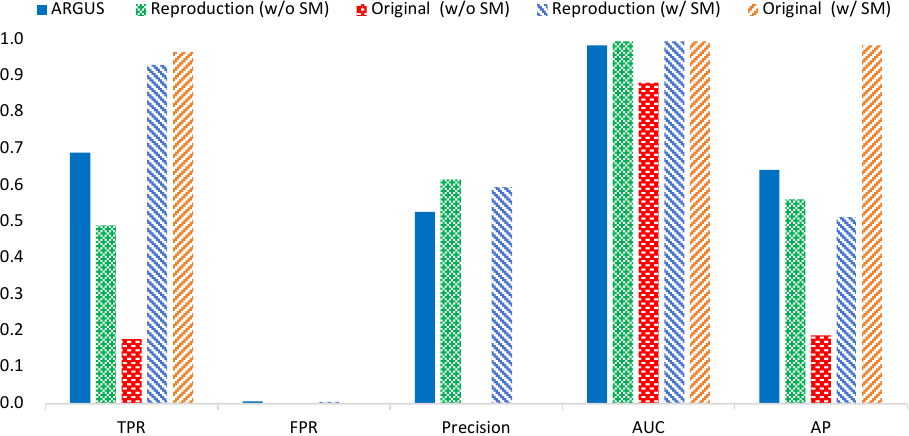}
    \caption{Comparison of \euler results on the DARPA OpTC dataset. ``\argus'' represents the reproduction result of \euler in the \argus paper. ``Reproduction'' and ``Original'' represent the results of our reproduction experiments and original paper, respectively. Original precision scores are not included in the figure since they are not provided in the original paper.}
    \label{fig:rq1-Euler}
\end{figure}

For the \euler model, we also evaluated the \euler-SM LSTM version from the paper~\cite{king2023euler2}, which demonstrated the best detection performance. This model uses an LSTM network to learn the temporal features between dynamic temporal graphs and includes an additional softmax layer (SM) at the end to aggregate the embeddings of neighbor nodes. However, when we reproduced this evaluation experiment as described in the paper, the AP score of the \euler model, whether using SM or not, could not reach the original paper's results. Furthermore, we compared it with the reproduction results of the \euler model provided in the \argus paper. As shown in Figure~\ref{fig:rq1-Euler}, these results are close to those of our reproduction experiment.

\section{Evaluation on the CIC-IDS-2017 Dataset}
\label{sec:cic-ids-2017-results}
Table~\ref{tab:rq1} presents the experimental results of R+R of \gnids on the CIC-IDS-2017 dataset. ``\argus\_ft'' represents the detection performance of the model when considering communication features between nodes, and ``\argus'' represents the detection performance without accounting for these features.

\begin{figure}
    \centering
    \includegraphics[width=0.9\linewidth]{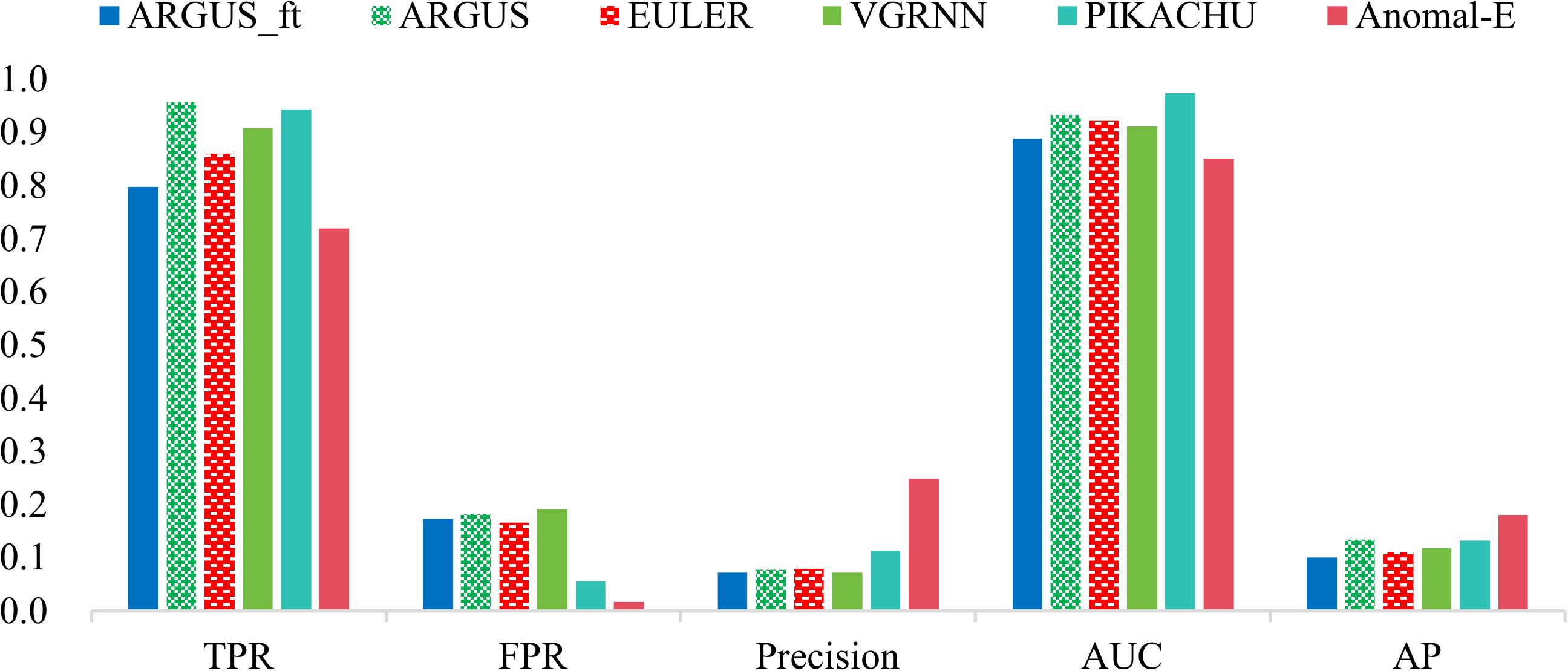}
    \caption{Evaluation on the small-scale dataset}
    \label{fig:rq2-9000}
\end{figure}

\section{Evaluation on the Small-scale New Dataset}
\label{sec:small-scale}
Figure~\ref{fig:rq2-9000} shows the results of \gnids on the small-scale dataset based on real-world enterprise traffic.

\section{Evasion Attack Performance of \euler \& \argus}
\label{sec:euler-argue-evasion-attack}
Tables~\ref{tab:rq5-euler} and~\ref{tab:rq5-argus} demonstrate the evasion attack performance of \euler \& \argus, respectively, on LANL, OpTC, and our own dataset.

\begin{table}[!ht]
\centering
    \caption{Evasion attack performance of EULER.}
    \label{tab:rq5-euler}
    \vspace{-2ex}
    \begin{tabular}{c|ccc|ccc|ccc}
    \hline
    \multirow{2}{*}{K} & \multicolumn{3}{c}{LANL} & \multicolumn{3}{c}{OpTC} & \multicolumn{3}{c}{Our Dataset} \\
    & $r_{tgt}$ & $r_{cov}$ & $r_{atk}$ & $r_{tgt}$ & $r_{cov}$ & $r_{atk}$ & $r_{tgt}$ & $r_{cov}$ & $r_{atk}$ \\ \hline
0 & 0.13 & 1.00 & 0.13 & 0.03 & 1.00 & 0.03 & 0.03 & 1.00 & 0.03 \\
2 & 1.00 & 1.00 & 1.00 & 0.38 & 1.00 & 0.38 & 0.03 & 0.99 & 0.00 \\
5 & 1.00 & 1.00 & 1.00 & 0.69 & 1.00 & 0.69 & 0.03 & 0.98 & 0.00 \\
10 & 1.00 & 1.00 & 1.00 & 1.00 & 1.00 & 1.00 & 0.03 & 0.98 & 0.00 \\
20 & 1.00 & 1.00 & 1.00 & 1.00 & 1.00 & 1.00 & 0.03 & 0.96 & 0.00 \\
50 & 1.00 & 1.00 & 1.00 & 1.00 & 1.00 & 1.00 & 0.03 & 0.94 & 0.00 \\ \hline
\end{tabular}
\end{table}

\begin{table}[!ht]
\centering
    \caption{Evasion attack performance of ARGUS.}
    \label{tab:rq5-argus}
    \vspace{-2ex}
    \begin{tabular}{c|ccc|ccc|ccc}
    \hline
    \multirow{2}{*}{K} & \multicolumn{3}{c}{LANL} & \multicolumn{3}{c}{OpTC} & \multicolumn{3}{c}{Our Dataset} \\
    & $r_{tgt}$ & $r_{cov}$ & $r_{atk}$ & $r_{tgt}$ & $r_{cov}$ & $r_{atk}$ & $r_{tgt}$ & $r_{cov}$ & $r_{atk}$ \\ \hline
0 & 0.18 & 1.00 & 0.18 & 0.02 & 1.00 & 0.02 & 0.04 & 1.00 & 0.04 \\
2 & 1.00 & 1.00 & 1.00 & 0.20 & 1.00 & 0.20 & 0.04 & 0.98 & 0.03 \\
5 & 1.00 & 1.00 & 1.00 & 0.50 & 1.00 & 0.50 & 0.04 & 0.97 & 0.03 \\
10 & 1.00 & 1.00 & 1.00 & 0.98 & 1.00 & 0.98 & 0.04 & 0.97 & 0.03 \\
20 & 1.00 & 1.00 & 1.00 & 1.00 & 1.00 & 1.00 & 0.13 & 0.99 & 0.13 \\
50 & 1.00 & 1.00 & 0.99 & 1.00 & 1.00 & 1.00 & 0.22 & 1.00 & 0.21  \\ \hline
\end{tabular}
%\vspace{-2ex}
\end{table}